\documentclass[journal,comsoc]{IEEEtran}
\usepackage{cite}
\usepackage{bm}
\usepackage{amsmath}
\usepackage{extarrows}
\usepackage{amssymb}
\usepackage{graphicx}
\usepackage{color}
\usepackage{enumerate}
\usepackage{bookmark} 
\graphicspath{{figure}}
\usepackage{setspace}
\usepackage{subfigure}
\usepackage{algorithm}
\usepackage{algorithmicx}
\usepackage{multirow}
\usepackage{stfloats}

\newcommand{\mr}{\mathrm} 

\newcommand{\BE}{\begin{equation}}
\newcommand{\EE}{\end{equation}}
\newcommand{\BS}{\begin{subequations}}
\newcommand{\ES}{\end{subequations}}
\renewcommand{\bf}{\bm}

\newtheorem{theorem}{Theorem}

\newtheorem{property}{Property}

\newtheorem{lemma}{Lemma}

\newcommand{\tabincell}[2]{\begin{tabular}{@{}#1@{}}#2\end{tabular}}

\allowdisplaybreaks \allowdisplaybreaks[2]

\ifCLASSINFOpdf
\graphicspath{{figure/}}

\else

\fi
\begin{document}

\title{{Constrained Capacity Optimal Generalized Multi-User MIMO: A Theoretical and Practical Framework}}
\author{{\IEEEauthorblockN{Yuhao Chi, \emph{Member, IEEE}, Lei Liu*,  \emph{Member, IEEE}, Guanghui Song, \emph{Member, IEEE},\\ Ying Li,  \emph{Member, IEEE}, Yong Liang Guan, \emph{Senior Member, IEEE}, and Chau Yuen, \emph{Fellow, IEEE}}}
	
\thanks{Yuhao Chi, Guanghui Song, and Ying Li are with the State Key Lab of Integrated Services Networks, Xidian University, Xi'an, 710071, China (e-mail: yhchi@xidian.edu.cn, gsong2017@gmail.com, yli@mail.xidian.edu.cn).}
\thanks{Lei Liu is with the School of Information Science, Japan Advanced Institute of Science and Technology (JAIST), Nomi 923-1292, Japan (*Corresponding author, e-mail: leiliu@jaist.ac.jp).}
\thanks{Yong Liang Guan is with the School of Electrical and Electronic Engineering, Nanyang Technological University, Singapore 639798 (e-mail: eylguan@ntu.edu.sg).}
\thanks{Chau Yuen is with the Singapore University of Technology and Design, Singapore 487372 (e-mail: yuenchau@sutd.edu.sg).}
\thanks{This article has been presented in part at the 2022 IEEE ISIT, Finland, \cite{YuhaoISIT2022}.}
}

\maketitle
\begin{abstract}
Conventional multi-user multiple-input multiple-output (MU-MIMO) mainly focused on Gaussian signaling, independent and identically distributed (IID) channels, and a limited number of users. It will be laborious to cope with the heterogeneous requirements in next-generation wireless communications, such as various transmission data, complicated communication scenarios, and unprecedented massive user access. Therefore, this paper studies a generalized MU-MIMO (GMU-MIMO) system with more generalized and practical constraints, i.e., practical channel coding, non-Gaussian signaling, right-unitarily-invariant channels (covering Rayleigh fading channel matrices, certain ill-conditioned and correlated channel matrices, etc.), and massive users and antennas. These generalized assumptions bring new challenges in theory and practice. For example, there is no accurate constrained capacity region analysis for GMU-MIMO. In addition, it is unclear how to achieve constrained-capacity-optimal performance with practical complexity. 
	
To address these challenges, a unified framework is proposed to derive the constrained capacity region of GMU-MIMO and design a constrained-capacity-optimal transceiver, which jointly considers encoding, modulation, detection, and decoding. Group asymmetry is developed to group users according to their rates, which makes a tradeoff between user rate allocation and implementation complexity. Specifically, the constrained capacity region of group-asymmetric GMU-MIMO is characterized by using the minimum mean-square error (MMSE) optimality of orthogonal/vector  approximate message passing (OAMP/VAMP) and the relationship between mutual information and MMSE. Furthermore, a theoretically optimal multi-user OAMP/VAMP receiver and practical multi-user low-density parity-check (MU-LDPC) codes are proposed to achieve the constrained capacity region of group-asymmetric GMU-MIMO. Numerical results demonstrate that the proposed MU-LDPC coded GMU-MIMO systems achieve asymptotic performance within $0.2$ dB from the theoretical sum capacity. Moreover, their finite-length performances are about 1$\sim$2~dB away from the associated sum capacity of GMU-MIMO. 
\end{abstract}


\section{Introduction}
With the rapid development of wireless communications, a variety of communication services have emerged, such as heterogeneous vehicular networks~\cite{Ekram2021TMC}, in which the number of wireless-enabled devices is predicted to reach $10$ million connections per square kilometer in 6G\cite{Ding2021IeeeCM}. To support large-scale connectivity, multi-user multiple-input multiple-output (MU-MIMO)\cite{Rusek2013,GoutayJSAC2021,McDonald,Lozano2006TIT,WeiYuTIT2004, XiaojunTIT2014,LeiTSP2019,YuhaoTWC2018,LeiTWC2019,LeiTWC2016} is a popular technology that can provide lots of space resources for massive data transmissions. However, due to various data, complex channels, and large numbers of users and antennas, the communication scenarios become more complex, which brings many new challenges to MU-MIMO not only in theory but also in practice.
 
\subsection{Information-Theoretical Limit of MU-MIMO}
A common information-theoretical limit that is frequently used in communication systems is channel capacity. It is well known that capacity is defined by default as the maximum mutual information over all possible choices of the input distribution. We employ a constrained capacity for GMU-MIMO, which is defined as mutual information under a fixed input distribution, due to the arbitrarily fixed input distribution constraint.  It should be emphasized that the Gaussian capacity, i.e. Shannon capacity in the strict sense for additive white Gaussian noise (AWGN) channel, is a special case of the constrained capacity when the signaling is Gaussian. 

For available channel state information (CSI) at the transceiver, a kind of water-filling technique was proposed to obtain the Gaussian capacity for Gaussian signaling\cite{McDonald,WeiYuTIT2004} and constrained capacity of MU-MIMO  with arbitrarily distributed input signaling\cite{Lozano2006TIT}. When CSI is only available at the receiver, the capacity of Gaussian MU-MIMO was approached with Gaussian signaling~\cite{tse2005fundamentals}. For arbitrarily distributed input signaling, only the constrained capacities of point-to-point MIMO (P2P-MIMO) were derived with independent and identically distributed (IID) channel matrices~\cite{Barbier2018b,ReevesTIT2019} or right-unitarily-invariant channel matrices~\cite{Barbier2017arxiv, LeiOptOAMP}, leveraging random matrix theory~\cite{Barbier2018b, Barbier2017arxiv, ReevesTIT2019} and approximate message passing (AMP)-type algorithms~\cite{LeiOptOAMP}, respectively. However, the constrained capacity region of MU-MIMO with arbitrarily distributed input signaling and right-unitarily-invariant channel matrices is still unknown.

\subsection{Practical Information-Theoretically Optimal Receivers of MU-MIMO}
It was proved that the capacity region of MU-MIMO can be achieved by a successive interference cancellation (SIC) receiver with time-sharing technology\cite{InforTh}. However, the following inherent problems make the SIC receiver impractical in large-scale systems: 1) severe decoding delay, 2) excessive overhead, 3) serious error propagation, and 4) prohibited complexity of optimizations in user decoding order and grouping strategy. To solve the issues of the SIC receiver, a lot of literature focused on parallel interference cancellation (PIC) receivers for MU-MIMO\cite{LeiTSP2019,YuhaoTWC2018,LeiTWC2016,LeiTWC2019}. 

For CSI only available at the receiver, when properly designed forward error correction (FEC) codes, optimality of the iterative linear minimum mean-square error (Turbo-LMMSE) receiver was proved to achieve the sum capacity of MU-MIMO with Gaussian signaling\cite{LeiTSP2019,YuhaoTWC2018}. To reduce the implementation complexity of LMMSE, Gaussian message passing receivers were proposed in\cite{LeiTWC2016,LeiTWC2019}. Nevertheless, Gaussian signaling is only an ideal concept. In practice, non-Gaussian discrete signaling is generally used, such as quadrature phase-shift keying (QPSK) and quadrature amplitude modulation (QAM). In this case, these Turbo-type receivers~\cite{LeiTSP2019,YuhaoTWC2018,LeiTWC2016,LeiTWC2019} are not capacity optimal anymore~\cite{LeiOptOAMP, MaTWC2019, LeiTIT2021}.

To address the above issue, AMP with well-designed FEC codes can achieve the constrained capacity of P2P-MIMO with IID channel matrices and arbitrary input signaling \cite{LeiTIT2021}. Moreover, for discrete signaling, the achievable rate of AMP was shown to be higher than that of Turbo-LMMSE. However, AMP is limited to IID channel matrices \cite{MaTWC2019}. For non-IID channels, AMP performs poorly or even diverges \cite{Vila2015ICASSP,manoel2014sparse,Rangan2017TIT}, such that the results in \cite{LeiTIT2021} will become invalid. 

To overcome the limitation of AMP on non-IID channels, orthogonal AMP (OAMP)\cite{MaAcess2017} and vector AMP (VAMP)\cite{Rangan2019TIT} were proposed to offer improved performances in a wide range of right-unitarily-invariant matrices. Compared with AMP, OAMP and VAMP can be applied to more complex and practical communication scenarios, i.e., right-unitarily-invariant matrices, covering certain ill-conditioned and correlated
channel matrices\cite{MaTWC2019, Poor2021TWC}. In\cite{LeiOptOAMP}, it provided the rigorously capacity optimality proof for OAMP in P2P coded linear (i.e., MIMO) systems with right-unitarily-invariant matrices and arbitrary input distributions. Furthermore, it was shown that OAMP outperforms AMP in un-coded physical random access channels\cite{LeiJSAC2021}, generalized frequency division multiplexing (GFDM)\cite{JinSPAWC2017}, extra-large-scale massive MIMO\cite{Jintwc2020}, and the conventional Turbo receivers in P2P-MIMO channels\cite{MaTWC2019, LeiOptOAMP}.
Note that the optimal coding design is not investigated in works\cite{LeiJSAC2021,JinSPAWC2017,Jintwc2020}, making it difficult for OAMP to achieve constrained-capacity-optimal performance. Moreover,
due to the equivalence of OAMP and VAMP, they are referred to as OAMP/VAMP in this paper.

In a nutshell, current results are restricted to Gaussian signaling\cite{LeiTWC2016,LeiTSP2019,LeiTWC2019,YuhaoTWC2018}, IID channels\cite{LeiTIT2021}, un-coded systems\cite{LeiJSAC2021,JinSPAWC2017,Jintwc2020}, or P2P channels\cite{MaTWC2019, LeiOptOAMP}.

\subsection{Contributions of This Paper}
In next-generation wireless communications, due to the heterogeneous requirements such as diverse data transmission, complex communication scenarios, and unprecedented massive user access, the idealized assumptions of MU-MIMO (e.g., Gaussian signaling, IID channel matrices, a limited number of users and antennas, or CSI available at the transceiver) are difficult to hold. Therefore, this paper considers a generalized MU-MIMO (GMU-MIMO) with the more generalized and practical assumptions: 1) practical channel coding, 2) arbitrary input distributions, 3) general right-unitarily-invariant channel matrices, including Rayleigh fading matrices, certain ill-conditioned and correlated matrices\cite{MaTWC2019, Poor2021TWC}, 4) massive users and antennas, and 5) CSI only available at the receiver. 
However, the information-theoretical limit (i.e., constrained capacity region) and the low-complexity information-theoretically optimal receiver of GMU-MIMO are still open issues.  

To address the above challenges, we propose a unified framework to accurately characterize the constrained capacity region and design a constrained-capacity-optimal transceiver of GMU-MIMO. First of all, to meet the different rate requirements of users, group asymmetry is developed to achieve a good tradeoff between implementation complexity and rate allocation. That is, all users are divided into groups, and users with the same rate belong to one group. Secondly, since it is NP-hard to characterize the constrained capacity region of GMU-MIMO by straightforwardly calculating mutual information,  we derive the constrained capacity region of group-asymmetric GMU-MIMO, leveraging the MMSE optimality of OAMP/VAMP\cite{Barbier2017arxiv,Kabashima2006} and the relationship between mutual information and MMSE (I-MMSE)\cite{GuoTIT2005}. Specifically, the area covered by the MMSE transfer curves of OAMP/VAMP equals the constrained sum capacity of GMU-MIMO, based on which the constrained capacity region of group-asymmetric GMU-MIMO is derived. Then, we propose a practical multi-user OAMP/VAMP (MU-OAMP/VAMP) receiver for GMU-MIMO. An optimal design principle of multi-user codes is presented for MU-OAMP/VAMP to achieve the constrained capacity region of group-asymmetric GMU-MIMO. Moreover, a kind of multi-user low-density parity-check (MU-LDPC) code is designed for MU-OAMP/VAMP.

The major contributions of this paper are summarized as follows.

\begin{enumerate}
	\item Group-asymmetric GMU-MIMO is developed and its constrained capacity region is characterized accurately.
	
	\item The achievable sum rate and constrained-sum-capacity optimality of MU-OAMP/VAMP are analyzed and proved, based on which the optimal design principle of multi-user codes is presented for GMU-MIMO.
	
	\item A kind of capacity-approaching MU-LDPC code is designed for MU-OAMP/VAMP, whose theoretical detection thresholds are about 0.2 dB away from the constrained sum capacity.
	
	\item Numerical results show that the finite-length performances of the proposed framework with optimized MU-LDPC codes and QPSK modulation are about 1$\sim$2 dB from the associated sum capacity. They also outperform the existing state-of-art methods such as Turbo-LMMSE with optimized LDPC codes, and OAMP/VAMP with well-designed irregular P2P-LDPC codes. 
\end{enumerate}

In summary, this is the first work to provide a constrained-capacity-optimal framework for GMU-MIMO with practical complexity,  which is also the first time to apply the  OAMP/VAMP to GMU-MIMO. The main theoretical results of this paper, such as the characterization of the constrained capacity region, the optimal design principle of multi-user codes, and the constrained-capacity optimality proof of MU-OAMP/VAMP, are available for all signal constellations.

Part of the results in this paper has been published in~\cite{YuhaoISIT2022}. In this paper, we additionally provide the derivation of the constrained capacity region, detailed proofs, and more numerical results.

\begin{table*}[!t] \footnotesize
	\caption{Overview of the Closely Related Existing Works }\label{Overview}
	\centering\setlength{\tabcolsep}{0.5mm}{
		\begin{tabular}{||c|c|c|c|c|c|c||}
			\hline
			\multicolumn{3}{||c|}{System Model}       & \multirow{2}{*}{~Capacity  type~} &  \multirow{2}{*}{~Algorithm ~} & \multirow{2}{*}{{Algorithm optimality}} & \multirow{2}{*}{~Coding scheme~}\\ \cline{1-3}
			~P2P/MU/GMU-MIMO~ & ~Signaling~ & ~Channel Matrix~ &  &  &  &\\ 
			\hline
			\multirow{4}{*}{\centering P2P-MIMO} & \multirow{4}{*}{Arbitrary} & IID &\tabincell{c}{Constrained \vspace{-0.1cm}\\ \cite{ReevesTIT2019,Barbier2018b, LeiTIT2021}}& ~AMP\cite{donoho2009message,BayatiTIT2011}~&\tabincell{c}{Bayes  optimal \cite{ReevesTIT2019,Barbier2018b} \vspace{-0.1cm}\\ Capacity  optimal \cite{LeiTIT2021}}& \tabincell{c}{Uncoded \cite{ReevesTIT2019,Barbier2018b} \vspace{-0.1cm}\\P2P code~\cite{LeiTIT2021}}\\  
			\cline{3-7}
			  &  & \tabincell{c}{right-unitarily- \vspace{-0.1cm}\\invariant} &\tabincell{c}{Constrained \vspace{-0.1cm}\\\cite{Barbier2017arxiv, LeiOptOAMP}}& \tabincell{c}{~OAMP/VAMP \cite{MaAcess2017}, \vspace{-0.1cm}\\ VAMP \cite{Rangan2019TIT}, \vspace{-0.1cm}\\ EP \cite{Minka2013,opper2005expectation} }  &\tabincell{c}{ Bayes  optimal \cite{ Kabashima2006, Barbier2017arxiv} \vspace{-0.1cm}\\ Capacity  optimal\cite{LeiOptOAMP} }& \tabincell{c}{Uncoded \cite{ Kabashima2006, Barbier2017arxiv} \vspace{-0.1cm}\\ P2P code~\cite{LeiOptOAMP} }\\ 
			\hline
			MU-MIMO  & Gaussian  & \tabincell{c}{right-unitarily- \vspace{-0.1cm}\\invariant} &~Gaussian \cite{InforTh}~&Turbo-LMMSE\cite{XiaojunTIT2014}~ &\tabincell{c}{Capacity optimal\vspace{-0.1cm}\\\cite{LeiTSP2019,YuhaoTWC2018}}&~\tabincell{c}{{{Symmetric MU code~\cite{YuhaoTWC2018}}} \vspace{-0.1cm}\\ {{Asymmetric MU code~\cite{LeiTSP2019}}}}~\\ 
			\hline
			\textbf{GMU-MIMO} & \textbf{Arbitrary} & \tabincell{c}{\textbf{right-unitarily-} \vspace{-0.1cm}\\\textbf{invariant}} &~\tabincell{c}{\textbf{Constrained} \vspace{-0.1cm}\\ \textbf{[this paper]}} ~& \tabincell{c}{\textbf{MU-OAMP/VAMP} \vspace{-0.1cm}\\ 
				\textbf{[this paper]} } &~\tabincell{c}{\textbf{Capacity optimal} \vspace{-0.1cm}\\ \textbf{[this paper]}} ~ &~
			\tabincell{c}{{{\textbf{Asymmetric} \textbf{MU code}}} \vspace{-0.1cm}\\ \textbf{[this paper]}}~\\
			\hline 
	\end{tabular}} 
\end{table*}
\subsection{Connection to Existing Works}

\subsubsection{Relationship with EP} 
In fact, OAMP/VAMP is algorithmically equivalent to expectation propagation (EP)\cite{Minka2013,opper2005expectation, KeigoTIT2020}. For simplicity, we focus on OAMP/VAMP in this paper.  

\subsubsection{Other related low-complexity AMP-type algorithms} 
Recently, to avoid the high complexity LMMSE in OAMP/VAMP, low-complexity Bayes-optimal convolutional AMP (CAMP) \cite{Takeuchi2020CAMP}, memory AMP (MAMP) \cite{LeiMAMP}, and generalized MAMP (GMAMP) \cite{Lei_GMAMP} were proposed for right-unitarily-invariant matrices with arbitrary input distributions. Therefore, they may be good candidates with lower complexity for the proposed framework in this paper.

\subsubsection{Differences from Turbo-based MU-MIMO \texorpdfstring{\cite{LeiTWC2019,LeiTWC2016,LeiTSP2019,YuhaoTWC2018}}{}} Although there is a vast amount of works on MU-MIMO, most of their receivers are based on the conventional Turbo receiver~\cite{LeiTWC2019,LeiTWC2016,LeiTSP2019,YuhaoTWC2018}, which has been proven to be constrained-capacity-optimal for MU-MIMO systems with Gaussian signaling~\cite{LeiTSP2019}. In this paper, we consider GMU-MIMO with arbitrarily distributed signaling such as Gaussian, QPSK, QAM, etc. We will show that for non-Gaussian signaling, the Turbo-LMMSE is rigorously sub-optimal and worse than the proposed OAMP/VAMP receiver in Section~\ref{Any:Comping_Turbo}. Moreover, note that the coding rates of well-designed multi-user codes in \cite{LeiTSP2019,YuhaoTWC2018} are as low as $0.1$. In contrast, based on the OAMP/VAMP receiver, the proposed multi-user codes can support a larger rate range, such as $0.33 \sim 0.67$ in Section~\ref{sec:ach_max_sum}. See numerical results for more details.

\subsubsection{Differences from P2P-MIMO in \texorpdfstring{\cite{LeiTIT2021, LeiOptOAMP}}{}}
In \cite{LeiTIT2021, LeiOptOAMP}, AMP and OAMP were proven to be constrained-capacity-optimal for MIMO with arbitrary input signaling, but they are limited to P2P-MIMO that only involve one user with one rate. In contrast, This paper studies a more complicated GMU-MIMO that involves multiple users with different rates. As a result, the constrained-capacity region of GMU-MIMO needs to be established and the design of a constrained-capacity-optimal receiver becomes much more difficult. Therefore, the results in P2P-MIMO\cite{LeiTIT2021, LeiOptOAMP} cannot be straightforwardly applied to the constrained capacity region analysis and constrained-capacity-optimal receiver design for GMU-MIMO.

\subsubsection{Bayes optimality vs capacity optimality} MSE measurement is widely used in signal processing and detection of un-coded systems, which cannot characterize error-free recovery. As a result, MMSE is commonly used to denote the Bayes optimality (MMSE optimality) of an un-coded system. In contrast, the achievable rate is a key measurement for coded systems with error-free recovery.
Thus, the maximum achievable rate is commonly used to denote the capacity optimality of a coded system. It is worth noting that Bayesian optimality does not guarantee capacity optimality. This is demonstrated in the numerical results of this paper, where the performances of Bayes-optimal detectors with P2P capacity-approaching LDPC codes are rigorously sub-optimal. 

To show the difference between this paper and the existing closely related works, Table~\ref{Overview} presents the corresponding comparisons. In addition, a list of key abbreviations mainly used throughout the paper
is summarized in Table~\ref{acronyms}.

\begin{table*}[t]\scriptsize
	\centering
	\caption{List of key abbreviations.}\label{acronyms}
	\begin{tabular}{cc|ll|cc|ll}
		\hline
		\hline
		\multicolumn{2}{c|}{Abbreviations} & \multicolumn{2}{c|}{Definitions} & \multicolumn{2}{c|}{Abbreviations} & \multicolumn{2}{c}{Definitions} \\
		\hline
		\multicolumn{2}{c|}{MU-MIMO} & \multicolumn{2}{l|}{Multi-User Multiple-input Multiple-Output } & \multicolumn{2}{c|}{IID} & \multicolumn{2}{l}{Independent and Identically Distributed} \\
		\multicolumn{2}{c|}{GMU-MIMO} & \multicolumn{2}{l|}{Generalized Multi-User Multiple-input Multiple-Output } & \multicolumn{2}{c|}{LD} & \multicolumn{2}{l}{Linear Detection} \\
		\multicolumn{2}{c|}{MSE} & \multicolumn{2}{l|}{Mean Square Error} & \multicolumn{2}{c|}{NLD} & \multicolumn{2}{l}{Nonlinear Detection} \\
		\multicolumn{2}{c|}{MAP} & \multicolumn{2}{l|}{Maximum A Posteriori} & \multicolumn{2}{c|}{LDPC} & \multicolumn{2}{l}{Low-Density Parity-Check} \\
		\multicolumn{2}{c|}{MMSE} & \multicolumn{2}{l|}{Minimum Mean Square Error} & \multicolumn{2}{c|}{P2P} & \multicolumn{2}{l}{Point-to-Point} \\
		\multicolumn{2}{c|}{LMMSE} & \multicolumn{2}{l|}{Linear Minimum Mean Square Error} & \multicolumn{2}{c|}{APP} & \multicolumn{2}{l}{A-Posteriori Probability} \\
		\multicolumn{2}{c|}{AMP} & \multicolumn{2}{l|}{Approximate Message Passing} & \multicolumn{2}{c|}{SE} & \multicolumn{2}{l}{State Evolution} \\
		\multicolumn{2}{c|}{OAMP} & \multicolumn{2}{l|}{Orthogonal Approximate Message Passing} & \multicolumn{2}{c|}{QPSK} & \multicolumn{2}{l}{Quadrature Phase-Shift Keying} \\
		\multicolumn{2}{c|}{VAMP} & \multicolumn{2}{l|}{Vector Approximate Message Passing}& \multicolumn{2}{c|}{QAM} & \multicolumn{2}{l}{Quadrature Amplitude Modulation} \\
		\multicolumn{2}{c|}{MAMP} & \multicolumn{2}{l|}{Memory Approximate Message Passing} & \multicolumn{2}{c|}{SVD} & \multicolumn{2}{l}{Singular Value Decomposition} \\
		\hline
		\hline
	\end{tabular}
\end{table*}
\subsection{Notations}
Lowercase letters denote scalars and boldface lowercase letters denote vectors. $[\cdot]^{\rm{T}}$ and $[\cdot]^{\rm{H}}$ denote transpose and conjugate transpose operations respectively. $\mathbb{C}$ represents the complex field. $\bf{I}$ is the identity matrix. $I(\bf{x},\bf{y})$ denotes mutual information between $\bf{x}$ and $\bf{y}$. $|\mathcal{S}|$ is the cardinality of set $\mathcal{S}$. $\mr{Tr}(\bf{A})$ and ${\mr{det}}(\bf{A})$ for the trace and the determinant of $\bf{A}$. $\|\bf{a}\|$ for the $\ell_2$-norm the vector $\bf{a}$. $\mr{E}\{\cdot\}$ for the expectation over all random variables included in the brackets. $\mr{E}\{a|b\}$ for the conditional expectation of $a$ for given $b$. $\rm{mmse}\{a|b\}$ for $\mr{E}\{(a-E\{a|b\})^2|b\}$. $\mathcal{CN}(\bf{\mu},\bf{\Sigma})$ for the circularly-symmetric Gaussian distributions with mean $\bf{\mu}$ and covariance $\bf{\Sigma}$. The MMSE and constrained capacity of P2P-MIMO with IID Gaussian channel matrices\cite{ReevesTIT2019,Barbier2018b} or a sub-class of right-unitarily-invariant channel matrices\cite{Barbier2017arxiv} can be predicted by the replica method\cite{Kabashima2006,Tulino2013TIT,MaAcess2017}. For more types of channel matrices, the rigorous proof of the replica method is still an open issue. In this paper, we assume that the replica method is reliable and verify it with experimental results. For simplicity, MMSE and constrained capacity are not explicitly stated as the replica method in this paper.

\subsection{Paper Outline}
This paper is organized as follows. Section II presents the system model of GMU-MIMO. The constrained capacity region of group-asymmetric GMU-MIMO, achievable sum rate analysis of MU-OAMP/VAMP receiver, and the principle of multi-user code design are derived in Sections III and IV respectively. Section~V presents an example of two user-group GMU-MIMO. Numerical simulations are provided in Section VI and the conclusion is presented in Section~VII.

\section{System Model and Challenges}
In this section, the model and assumptions of GMU-MIMO systems are provided. Meanwhile, the key challenges of constrained capacity region analysis and multi-user code design are presented.

\subsection{System Model}\label{sec:system1}
\begin{figure*}[!t]
	\centering
	\includegraphics[width=0.75\linewidth]{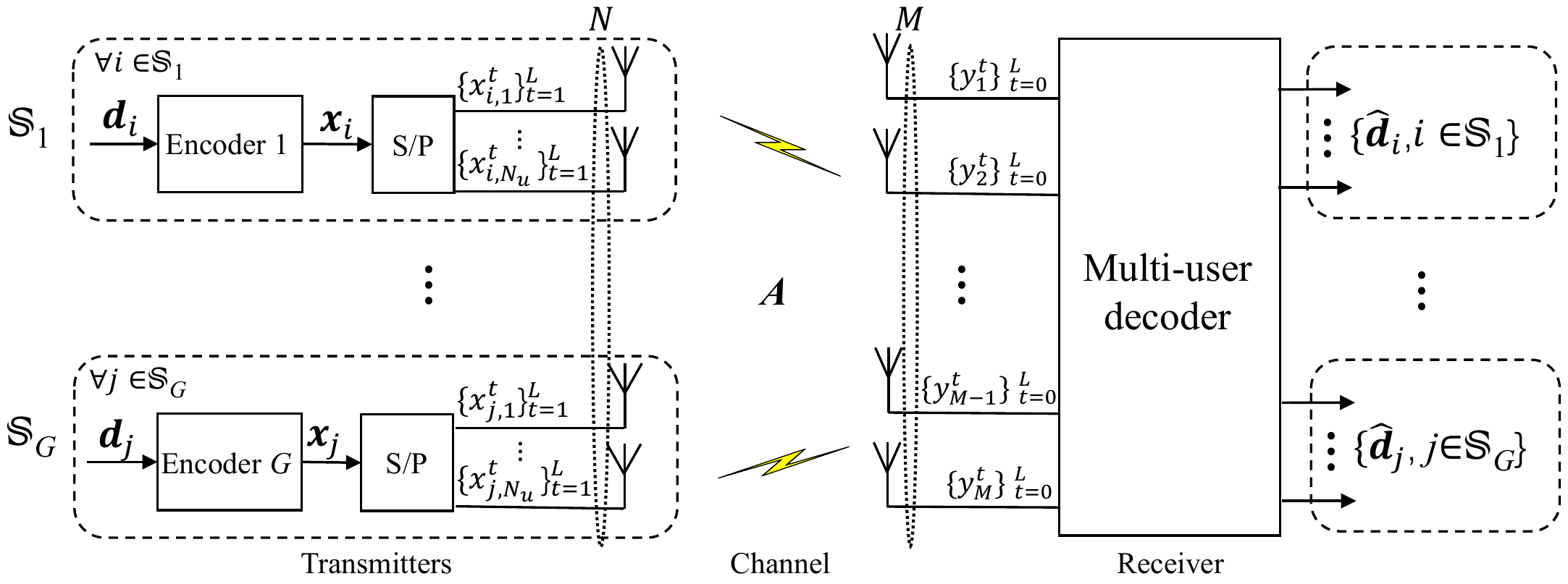}
	\caption{Illustration of a GMU-MIMO system with $K$ users partitioned into $G$ groups, and a receiver included a multi-user decoder. S/P denotes serial-to-parallel conversion. The number of transmitted and received antennas are $N$ and $M$ respectively.}\label{Fig:GMU-MIMO}
\end{figure*}
Fig.~\ref{Fig:GMU-MIMO} illustrates the uplink GMU-MIMO system with $K$ transmit users and one receiver. Total of $N$ transmit antennas are employed by the $K$ users and each user has $N_u=N/K$ antennas. The receiver has $M$ receive antennas. Users are equally partitioned into $G$ groups, where each group has $K/G$ users. Let  $\mathbb{S}_1, ...,\mathbb{S}_G$ be the sets that include the user indices of the $G$ groups, where $\mathbb{S}_i \cap \mathbb{S}_j = \emptyset$ for $i \ne j$, $i,j\in\mathbb{G}=\{1,...,G\}$, and $\mathbb{S}_1 \cup \mathbb{S}_2 \cdots \cup \mathbb{S}_{G} = \mathbb{K} \equiv \{1,\cdots, K\}$. Users in the same group employ the same encoder with the same code rate and that in different groups employ different encoders with different code rates. 

\emph{Note:} The proposed framework can be easily extended to the more general configurations, i.e., each group consists of a different number of users, and each user is equipped with a different amount of antennas. For more details, see Section~\ref{Sec:general_feasibility}.

At the transmission side, since the processing of each user's data is similar, we describe the transmission of user  $i\in\mathbb{S}_1$.  Message vector $\bf{d}_i$  is encoded by encoder 1 and the output codeword is denoted by  $\bf{x}_i$. We assume $\bf{x}_i$ is a modulated signal vector whose entries are from a constellation set $\cal{S}$.  A serial-to-parallel conversion (S/P) is employed to produce the transmit signals over each antenna. Suppose the length of $\bf{x}_i$ is $N_uL$ for a given integer $L$. Codeword $\bf{x}_i$ is split into $N_u$ length-$L$ vectors $\bf{x}_{i,n}, n=1,...,N_u$, and $\bf{x}_{i,n}=\{x^t_{i,n}\}_{t=1}^L$ is transmitted  over antenna $n$. $L$ is the total transmission time for codeword $\bf{x}_i$. At time $t$, the transmission signals of user $i$ over the $N_u$ antennas are $\bf{x}_i^t=(x^t_{i,1},...,x^t_{i,N_u})$. The all transmitted signals of $K$ users are denoted as $\bf{x}^t=(x^t_{1,1},..., x^t_{G,N_u}) \in \mathbb{C}^{N \times 1}$ which satisfy the power constraint $\tfrac{1}{N}\mr{E}\{\|\bf{x}^t\|^2\}=1$.

The receiver obtains signal $\bf{y}^t= [y_1^t, ..., y_M^t]^{\rm{T}}$, given by
\BE\label{Eqn:gmu_recv}
\bf{y}^t=\bf{A}\bf{x}^t+\bf{n}^t,\;\;  t=1,\dots,L,
\EE
where $\bf{A}\in \mathbb{C}^{M\times N}$ is a channel matrix and $\bf{n}^t \sim\!\mathcal{CN}(\mathbf{0},\sigma^2\bm{I})$ is an AWGN vector. Without loss of generality, we assume $\tfrac{1}{N}{\rm tr}\{\bf{A}^{\rm{H}}\bf{A}\}=1$ and the signal-to-noise ratio (SNR) is defined as  ${snr} = \sigma^{-2}$.
Based on $\bf{y}^t$, a multi-user decoder is employed to recover the $K$ users' messages.

\subsection{Assumptions and Challenges of GMU-MIMO}\label{sec:GMU-MIMO}

\subsubsection{Assumptions} The GMU-MIMO system satisfies the following assumptions. 
\begin{itemize}
	\item A large-scale system is considered including massive users, massive transmitted and received antennas, i.e., $N \to \infty$, $M \to \infty$, and channel load $\beta=\frac{N}{M}$ is fixed.
	
	\item The entries of signal $\bf{x}$ follow an arbitrary distribution\footnote{For Gaussian signaling, the information-theoretic limit of the system in \eqref{Eqn:gmu_recv} is the well-known Gaussian sum capacity $C_{\rm Gau}= \log |\bf{I} +  snr\bf{A}^H\bf{A}|$ \cite{tse2005fundamentals}, which can be achieved by Turbo-LMMSE \cite{LeiTSP2019}. However, for non-Gaussian signaling, the constrained capacity region of the system in \eqref{Eqn:gmu_recv} is not trivial, and the constrained-capacity-optimal transceiver with practical complexity remains an open issue.}  (e.g., BPSK, QPSK, QAM, Gaussian, Bernoulli-Gaussian, etc.). 

	\item Channel matrix $\bf{A}$ is right-unitarily-invariant, 
	  which covers a variety of fading channel models including the commonly used IID random (i.e., Rayleigh fading) matrices and certain ill-conditioned (e.g., correlated) matrices~\cite{MaTWC2019,Poor2021TWC}. That is, let the SVD of $\bf{A}$ be $\bf{A} = \bf{U}\bf{\Lambda}\bf{V}$, where $\bf{U} \in \mathbb{C}^{M\times M}$ and $\bf{V} \in \mathbb{C}^{N\times N}$ are unitary matrices, and $\bf{\Lambda}$ is a rectangular diagonal matrix. $\bf{U}$, $\bf{V}$, and $\bf{\Lambda}$ are mutually independent, and $\bf{V}$ is Haar-distributed \cite{KeigoTIT2020}.
	
	\item Channel matrix $\bf{A}$ is available at the receiver but unknown at the transmitters\footnote{In large-scale MU-MIMO, CSI at the transmitters is impractical as it brings a huge overhead cost. When CSI is unavailable at transmitters, the conventional precoding and water-filling power allocation are unavailable.}. 
	
	\item The users are group-asymmetric\footnote{Due to the limitation of complexity, it is prohibited to design transceivers for a completely asymmetric system that all users may have different rates. Group asymmetry makes a good tradeoff between the system complexity and rate allocation.}, i.e., all users are partitioned into multiple groups, where user rates of the same group are the same and different groups may have different user rates. 
	
\end{itemize}
\subsubsection{Challenges} The above assumptions bring new challenges to conventional MU-MIMO technologies in theory and practice.
\begin{itemize}
\item For arbitrarily distributed $\bf{x}$ and right-unitarily-invariant $\bf{A}$, there is no accurate constrained capacity region analysis of the system in \eqref{Eqn:gmu_recv}. The reason is that the existing capacity analyses for conventional MU-MIMO mainly focus on Gaussian signaling\cite{LeiTSP2019} or IID channel matrices~\cite{LeiTIT2021}.
		
\item How to achieve the optimal performance of GMU-MIMO with practical complexity is still an open issue. The globally \emph{maximum a posteriori} (MAP) receiver is the optimal solution~\cite{verdu1984optimum}, but it is unusable for large-scale systems due to its prohibitive complexity. Due to arbitrarily distributed $\bf{x}$ and right-unitarily-invariant $\bf{A}$, the existing practical capacity optimal receivers such as Turbo-LMMSE~\cite{LeiTSP2019} and AMP~\cite{LeiTIT2021} are sub-optimal.
		
\item The design principle of practical multi-user codes for GMU-MIMO systems is still unclear, particularly for group-asymmetric cases. The conventional capacity-approaching P2P codes~\cite{ryan2009channel,Richardson2001,Kim2009} are designed specially to overcome channel noises, which ignores the impact of large-scale antennas and users. In~\cite{XiaojunTIT2014,LeiTIT2021,LeiOptOAMP}, the channel codes are designed for one user in P2P-MIMO, which can not apply to group-asymmetric GMU-MIMO involving different requirements of users. Moreover, the existing multi-user channel codes~\cite{LeiTWC2016,LeiTSP2019,LeiTWC2019,YuhaoTWC2018} are designed for the conventional Turbo receiver in MU-MIMO, which is optimal for Gaussian signaling and worse for discrete signaling. In other words, the design principle of channel codes  in~\cite{LeiTIT2021,LeiTWC2016,LeiTSP2019,LeiTWC2019,YuhaoTWC2018,XiaojunTIT2014,LeiOptOAMP} is only available for P2P-MIMO or MU-MIMO, which cannot be applied to GMU-MIMO.
\end{itemize}

\section{Multi-user OAMP/VAMP Receiver and Constrained Capacity Region Characterization of GMU-MIMO}
In this section, we first present the multi-user OAMP/VAMP (MU-OAMP/VAMP) receiver for GMU-MIMO. Then, the constrained sum capacity and group capacity region for GMU-MIMO are accurately expressed. 
\subsection{MU-OAMP/VAMP Receiver}

Since the detection process of \eqref{Eqn:gmu_recv} in each time slot is the same, we omit the time index $t$ in the rest of this paper for simplicity. 

\begin{figure*}[!t]
	\centering
    \subfigure[Iterative detection]{
    \begin{minipage}[t]{0.5\linewidth}\label{Fig:MU-OAMP/VAMP} 
    	\centering
    \includegraphics[width=1\linewidth]{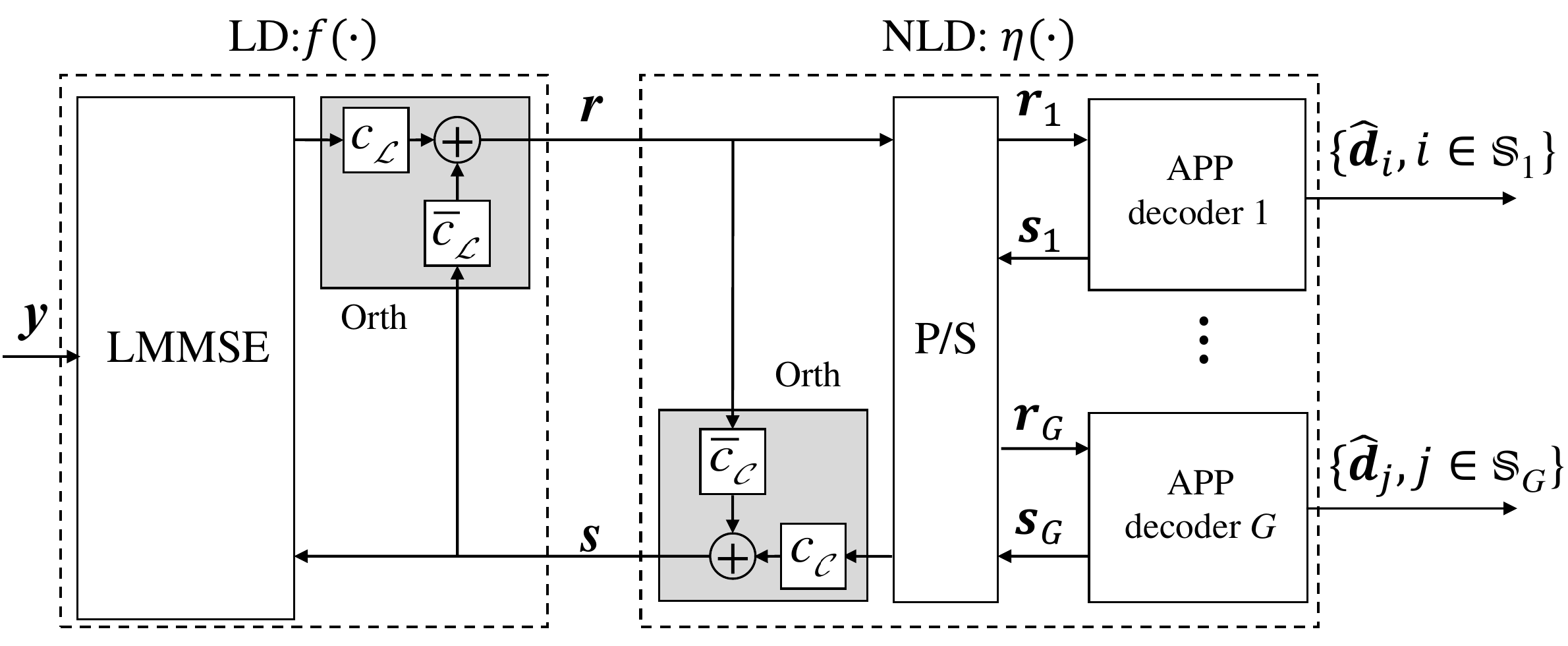}\vspace{0.1cm}
	\end{minipage}
	}
	\subfigure[Transfer functions]
	{
    \begin{minipage}[t]{0.5\linewidth}\label{Fig:coded_SE} 
    	\centering
    	\includegraphics[width=0.65\linewidth]{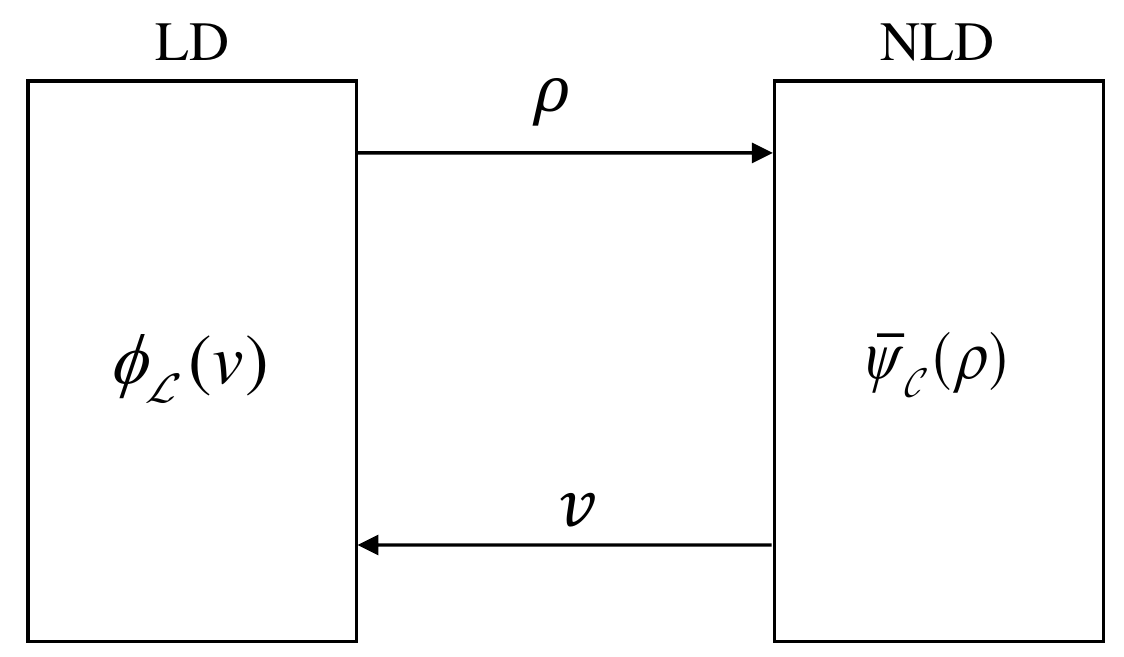}\vspace{0.4cm}
	\end{minipage}	
	}
	\caption{MU-OAMP/VAMP receiver consists of an LD and an NLD, where the LD employs LMMSE detection and the NLD consists of $G$ APP decoders. ``Orth'' represents the orthogonalization of the input and output  of LD and NLD.}
\end{figure*}
As shown in Fig.~\ref{Fig:MU-OAMP/VAMP}, MU-OAMP/VAMP consists of a linear detector (LD) $f(\cdot)$ and a non-linear detector (NLD) $\eta(\cdot)$, which employ LMMSE detection for linear constraint in~\eqref{Eqn:gmu_recv} and  \emph{a-posteriori probability}  (APP) decoding for code constraint $\bf{x} \in {\cal{C}}$ ($\cal{C}$ is the set of transmitted codewords), respectively. The NLD of MU-OAMP/VAMP is composed of a bank of APP decoders including demodulation and channel decoding. In addition, multiple iterations are performed between the LD and NLD, where the output estimations of LD are input to the NLD and then the updated output estimations of NLD are fed back to the LD.

The detailed process is given as follows: 
\BS\label{Eqn:MU-OAMP/VAMP}
\begin{align}
\mathrm{LD:}& \;\;\bf{r}= f(\bf{s})=  c_{\cal{L}} f_{\mr{lmmse}}({\bf{s}}) + {\bar{c}_{\cal{L}}} \bf{s},\label{Eqn:MULD}\\ 
\mathrm{NLD:}& \;\; \bf{s} = \eta(\bf{r})= c_{\cal{C}} \eta_{\mr{mmse}}({\bf{r}}) + {\bar{c}_{\cal{C}}} \bf{r},\label{Eqn:MUNLD}
\end{align}
\ES
where $\bar{c}_{\cal{L}}=1-c_{\cal{L}}$, $\bar{c}_{\cal{C}}=1-c_{\cal{C}}$, $\bf{r}=[r_1, ..., r_N]^{\mr{T}}$ and $\bf{s}=[s_1, ..., s_N]^{\mr{T}}$ denote the LD and NLD estimations of $\bf{x}$ respectively, and subscripts $\cal{L}$ and $\cal{C}$ indicate linear constraint and code constraint respectively. Let ${\bf{r}}_g$ and $\bf{s}_g$ be the input and output estimations of decoder~$g$ for the users in group $g$, $\forall g \in \mathbb{G}$.

In~\eqref{Eqn:MU-OAMP/VAMP}, the local estimation functions of LD and NLD are
\BS
\begin{align}
f_{\mr{lmmse}}({\bf{s}})&\equiv[{ snr }\bf{A}^{\rm{H}}\bf{A} +v_{s}^{-1}\bf{I}]^{-1}[{ snr }\bf{A}^{\rm{H}}\bf{y}+ v_{s}^{-1}{\bf{s}}],\label{Eqn:lmmse_Ax}\\ 
\eta_{\mr{mmse}}({\bf{r}})&\equiv \mr{E}\{ \bf{x} |\bf{r},\bf{x}\in {\cal{C}}\},\label{Eqn:mmse_Ck} 
\end{align}
\ES
where $ snr $ denotes the given SNR  and the calculation of \eqref{Eqn:mmse_Ck} consists of the demodulation and APP channel decoding \cite[Equation(10)]{MaTWC2019}. 
The parameters $c_{\mathcal{L}}$ and $c_{\mathcal{C}}$ in \eqref{Eqn:MU-OAMP/VAMP} are given by
\begin{align}
c_{\cal{L}}=\frac{v_{{s}}}{v_{{s}}- \Omega_{\cal{L}}(v_{s}^{-1})}  \quad {\rm and} \quad c_{\cal{C}} = \frac{v_{r}}{v_{{r}}- {\bar{\Omega}}_{\cal{C}}(\rho)}, 
\end{align}
where  $v_s$ and $v_r$ are the variances of $\bf{s}$ and ${\bf{r}}$ respectively, and the MMSE (per transmit antenna) functions are defined as
\BS\begin{align}
\Omega_{\cal{L}}(\rho)  &\equiv \tfrac{1}{N}{\mr{Tr}}\{[{ snr }\bf{A}^{\rm{H}}\bf{A}+\rho\bf{I}]^{-1}\},\label{Eqn:MU_rlmmse}\\
{\bar{\Omega}}_{\cal{C}}(\rho) & \equiv \tfrac{1}{G} \textstyle\sum\nolimits_{g=1}^{G}\Omega_{{\cal{C}}_g}(\rho), \\ \Omega_{{\cal{C}}_g}(\rho) &\equiv \tfrac{G}{N} \mr{E}\big\{\|\bf{x}_g-\eta_{\mr{mmse}}( {\sqrt{\rho}} \bf{x}_g + {\bf{z}})\|^2\big\},\label{Eqn:MU_slmmse}
\end{align} 
\ES
where $\bf{x}_g$ is the signal vector of the users in $\mathbb{S}_g$ and ${\bf{z}}\sim \mathcal{CN}(0,\bf{I})$ is independent of $\bf{x}_g$. 

All the asymmetric information is averaged during the iteration by virtue of the right-unitarily-invariant property of $\bf{A}$\cite{LeiOAMP2022}. As a consequence, the signals input to LD and NLD are effectively averaged, considerably simplifying the design of the MU-OAMP/VAMP receiver. Meanwhile, the orthogonalization in \eqref{Eqn:MU-OAMP/VAMP} is necessary to make the input-output estimated errors of LD and NLD uncorrelated during the iteration. The orthogonalization is denoted as ``Orth'' in Fig.~\ref{Fig:MU-OAMP/VAMP} and proved to ensure the exact state evolution of OAMP~\cite{MaAcess2017,MaTWC2019}. 

\emph{State evolution}: LD and NLD are exactly characterized by the state evolution (SE) as shown in Fig.~\ref{Fig:coded_SE}, in which the transfer functions are consisted of SNR function ${\bar{\psi}}_{\cal{C}}(\rho)$ and MSE function $\phi_{\cal{L}}(v)$ as follows.
\BS\label{Eqn:MUSE}
\begin{align}
\mathrm{LD:}&
\;\;\rho \equiv \phi_{\cal{L}}(\bar{v}) = \big[\tfrac{1}{N}\|\bf{r}-\bf{x}\|^2\big]^{-1},\label{Eqn:MUSE_LD}\\
\mathrm{NLD:}&\;\; \bar{v} \equiv {\bar{\psi}}_{\cal{C}}(\rho) \nonumber\\
&\quad= \tfrac{1}{G}\textstyle\sum_{g=1}^{G} \psi_{{\cal{C}}_g}(\rho) = \tfrac{1}{G}\sum_{g=1}^{G}\|\bf{s}_g-\bf{x}_g\|^2.\label{Eqn:MUSE_NLD}
\end{align}
\ES

The following lemma, proved in \cite{KeigoTIT2020,Rangan2019TIT,LeiOptOAMP}, shows the approximate IID Gaussianity of MU-OAMP/VAMP, which is critical to simplify the design and analysis of MU-OAMP/VAMP. 
\begin{lemma}[Approximate IID Gaussianity]\label{Lem:GA_MUSE}
	The input of NLD can be regarded as $\bf{r}  = \bf{x} + \rho^{-1/2}{\bf{z}}$, where ${\bf{z}}$ is an AWGN noise, ${\bf{z}}\sim \mathcal{CN}(\bf{0},\bf{I})$ independent of $\bf{x}$, and $\bf{x}$ is discrete signaling with $x_i \sim P_{\mathcal{S}}(x_i)$. As a result, (\ref{Eqn:MUSE_LD}) and (\ref{Eqn:MUSE_NLD}) are rewritten as 
	\BS\label{Eqn:MU-TF} \begin{align}
	\mathrm{LD:}& \quad   \rho =\phi_{\cal{L}}({\bar{v}})  = [{\Omega_{\cal{L}}({\bar{v}}^{-1})}]^{-1} -  {\bar{v}}^{-1},\label{Eqn:TFLD}\\
	\mathrm{NLD:}& \quad   \bar{v} = {\bar{\psi}}_{\cal{C}}(\rho) = \left([\bar{\Omega}_{\cal{C}}(\rho)]^{-1}- \rho \right)^{-1},\label{Eqn:TFNLD}
	\end{align}\ES
	where $\bar{v} \in[0,1]$ and $\phi_{\cal{L}}(0)=\tfrac{ snr }{N}{\rm tr}\{\bf{A}^{\rm{H}}\bf{A}\}={snr}$ due to the normalized singular values of $\bf{A}$, i.e., $\tfrac{1}{N}{\rm tr}\{\bf{A}^{\rm{H}}\bf{A}\}=1$.
\end{lemma}

Define $\phi_{\cal{L}}^{\mr{inv}}(\cdot)$ as the generalized inverse function of $\phi_{\cal{L}}(\cdot)$. For simplicity, let ${v}\equiv{\bar{v}}$. Based on \eqref{Eqn:MU-TF}, the variational transfer functions can be equivalently obtained as following.
\BS\label{Eqn:NewMUSE}
\begin{align}
\mathrm{LD:} \;\;   & {{v}} = \varphi_{\cal{L}}(\rho) \equiv \big(\rho + [\phi_{\cal{L}}^{\mr{inv}}(\rho)]^{-1} \big)^{-1}, \\
\mathrm{NLD:}  \;\; & {{v}} = {\bar{\Omega}}_{\cal{C}}(\rho).
\end{align}\ES

The I-MMSE lemma\cite{GuoTIT2005} will be used to derive the constrained capacity region of GMU-MIMO and the subsequent constrained-sum-capacity optimality proof of MU-OAMP/VAMP. The I-MMSE lemma, on the other hand, requires \emph{a-posteriori} estimations that do not generally satisfy the orthogonality requirements of MU-OAMP/VAMP. As a result, we employ the variational transfer function in \eqref{Eqn:NewMUSE}; for a more in-depth discussion, see\cite{LeiOptOAMP}. The variational transfer functions remain valid under the coding constraint, as demonstrated by simulations in\cite{MaTWC2019}.

\subsection{Constrained Sum Capacity of GMU-MIMO} 
For simplicity of discussions, we define an un-coded GMU-MIMO as
\BE\label{Eqn:cons_gmu_recv}
\bf{y}=\sqrt{\rho}\bf{A}\bf{x}+\bf{z},
\EE
where $\bf{z}\sim \mathcal{CN}(\bf{0}, \bf{I})$ is an AWGN noise vector, $\bf{x}=\{x_i\}$ is IID with $x_i\sim P_{\mathcal{S}}(x_i), \forall i$, and $\mathcal{S}$ denotes a constellation.  For convenience, we define
\BS\label{Eqn:Omega_S}\begin{align} 
\Omega_{\mathcal{S}}(\rho) &\equiv \tfrac{1}{N} \text{mmse} \big(\bf{x}|\sqrt{\rho}\bf{x}+\bf{z}, x_i\sim P_{\mathcal{S}}(x_i), \forall i\big),\\
\Omega_{x}(\rho) &\equiv \tfrac{1}{N} \text{mmse} \big(\bf{x}|\sqrt{\rho}\bf{A}\bf{x}+\bf{z}, x_i\sim P_{\mathcal{S}}(x_i), \forall i\big), \\
\Omega_{{A}x}(\rho) &\equiv \tfrac{1}{N} \text{mmse} \big(\bf{Ax}|\sqrt{\rho}\bf{A}\bf{x}+\bf{z}, x_i\sim P_{\mathcal{S}}(x_i), \forall i\big).
\end{align}\ES
Note that all the MMSE functions in this paper are defined on per transmit antenna.

Following the I-MMSE lemma~\cite{GuoTIT2005}, the average constrained capacity of GMU-MIMO per transmit antenna can be calculated by 
\BE\label{Eqn:const_C}
\bar{C} = \frac{1}{N}I(\bf{x}; \sqrt{snr}\bf{A}\bf{x}+\bf{z}) =\int_0^{snr} \Omega_{{Ax}}(\rho)\:d\rho,
\EE
which is reduced to $\int_0^{snr} \Omega_{{x}}(\rho)\:d\rho$ for P2P channels. 
Then, the constrained sum capacity of GMU-MIMO is 
\BE\label{Eqn:const_SC}
C_{\mr{GMU-MIMO}}^{\mr{sum}} = N \bar{C}.
\EE

Next, we derive the expression of measurement MMSE  $\Omega_{{Ax}}(\rho)$ in~\eqref{Eqn:const_C} using the properties of MU-OAMP/VAMP.

\begin{figure}[!t]
	\centering
	\includegraphics[width=0.7\columnwidth]{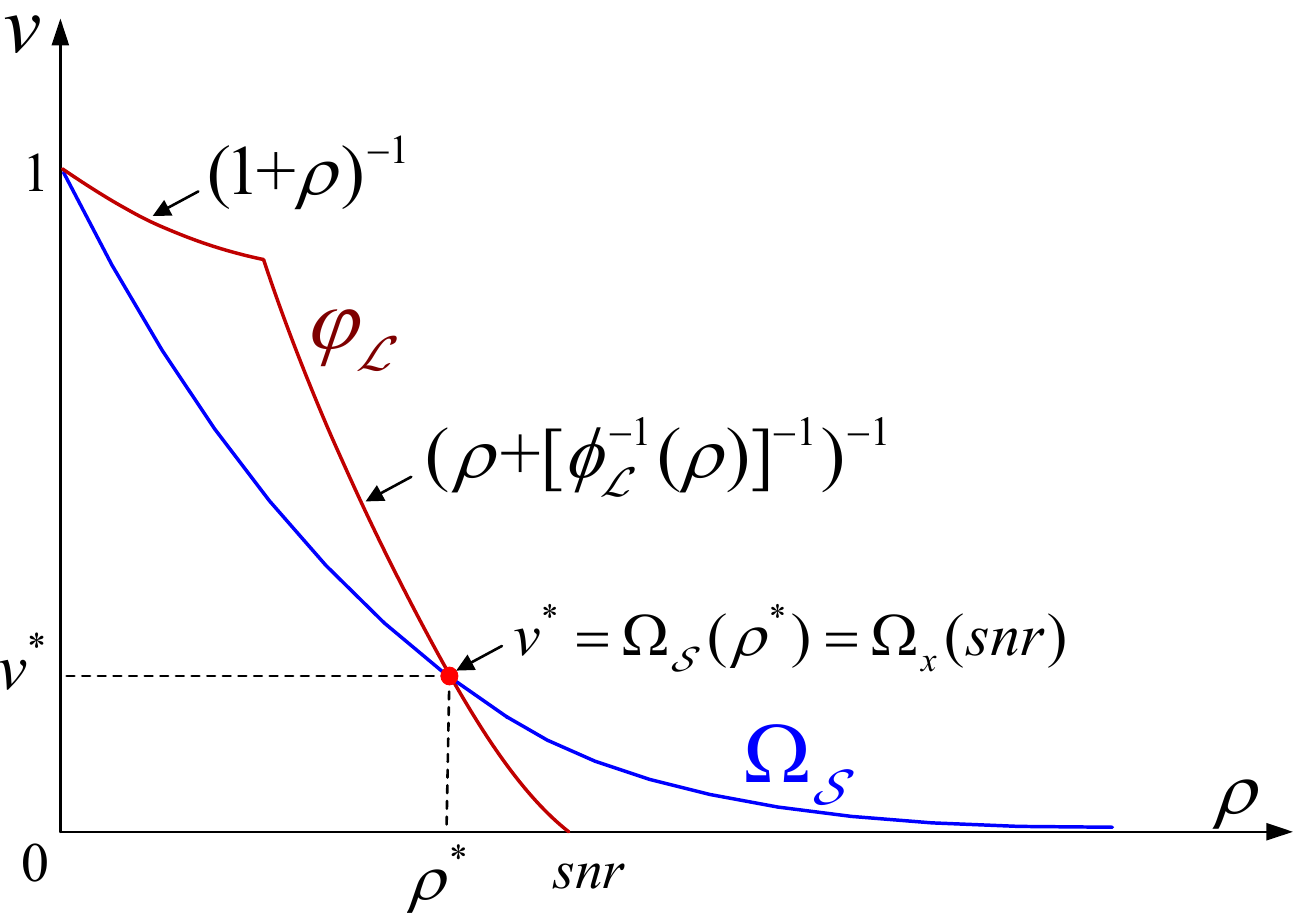} \\ 
	\caption{Illustration for transfer functions of MU-OAMP/VAMP in GMU-MIMO, where $\Omega_{\cal{S}}$ denotes the MMSE of constellation constraint and $(\rho^*,v^*)$ denotes the fixed point between $\Omega_{\cal{S}}$ and $\phi_{\cal{L}}$.}\label{Fig:un-coded_chart}
\end{figure}

As shown in Fig.~\ref{Fig:un-coded_chart}, the iterative process between LD $\phi_{\cal{L}}(\rho)$ and NLD  $\Omega_{\cal{S}}(\rho)$ converges to a unique fixed point $(\rho^*, v^*)$, where $v^*=\big([\Omega_{\mathcal{S}} (\rho^*)]^{-1} - \rho^*\big)^{-1}$ based on \eqref{Eqn:TFNLD}. The curve $\varphi_{\cal{L}}(\rho)$ is an upper bound of $ \Omega_{\cal{S}}(\rho)$ for $0\le \rho \le \rho^*$. According to the iterative process, the following lemma proved in \cite[APPENDIX D]{LeiOptOAMP} is given as follows.

\begin{lemma}[Measurement MMSE]\label{Lem:MMSE_Ax}
For the fixed point $(\rho^*, v^*)$, $\Omega_{Ax}(snr)$ is given by
\BS\label{Eqn:mmse_Ax}
\begin{align}
\Omega_{Ax}({ snr })&= \big(1\!-\!v^{*^{-1}}\Omega_{\mathcal{L}}(v^{*^{-1}})\big)/{ snr }\!=\! \rho^*\Omega_{\mathcal{S}}(\rho^*)/{ snr }\\
&= [1\!-\!v^{*^{-1}}\Omega_{x}({{ snr }})]/{ snr }\!=\!\rho^*\Omega_{x}({{ snr }})/{ snr }. 
\end{align}\ES 
\end{lemma}
Based on Lemma~\ref{Lem:MMSE_Ax}, \eqref{Eqn:const_C}, and \eqref{Eqn:const_SC}, the constrained sum capacity of GMU-MIMO in \eqref{Eqn:cons_gmu_recv} is given in the following theorem, which is proved based on the proof of~\eqref{Eqn:const_C} in \cite[APPENDIX D]{LeiOptOAMP}.

\begin{theorem}[Constrained Sum Capacity]\label{Them:Constrain_sum}
Assume that $\phi_{\cal{L}}(\rho)$ and  $\Omega_{\cal{S}}(\rho)$ has a unique fixed point $(\rho^*, v^*)$. Then, the constrained sum capacity of GMU-MIMO is given by
\BE\label{Eqn:const_SC_Express}
\!\!\!\!\!C_{\mr{GMU-MIMO}}^{\mr{sum}}\!= \!{\log \left| \bf{B}(v^*) \right|}+N \Big(\!\log  \Omega_{\mathcal{S}}(\rho^*)+\!\int_{0}^{\rho^*}\!\!\!\!{\Omega_{\mathcal{S}}(\rho) d\rho}\Big),
\EE
where $\bf{B}(v) = v^{-1}\bf{I} + { snr }\bf{A}^{\rm{H}}\bf{A}$. 
\end{theorem}

It is worth pointing out that for Gaussian signaling, the constrained sum capacity in~\eqref{Eqn:const_SC_Express} is degraded to the Gaussian sum capacity, i.e., $C_{\mr{GMU-MIMO}}^{\mr{sum}}=\!{\log \left|v^{-1}\bf{I} + { snr }\bf{A}^{\rm{H}}\bf{A} \right|}$~\cite{tse2005fundamentals}, where the MMSE function of Gaussian signaling is $\Omega_{\mr{Gau}}(\rho)=1/(1+\rho)$~\cite{Lozano2006TIT}.

\subsection{Constrained-Group-Capacity Region of GMU-MIMO}
For convenience, let 
\BS\label{Eqn:group_capacity_region}\begin{align} 
\Omega_{\mathcal{L}_{\mathbb{S}_\mathbb{Q}}}(v) & = \tfrac{1}{|{\mathbb{S}_\mathbb{Q}}|}{\mr {Tr}}\{[{ snr }\bf{A}_{\mathbb{S}_\mathbb{Q}}^{\rm{H}}\bf{A}_{\mathbb{S}_\mathbb{Q}}+v^{-1}\bf{I}_{\mathbb{S}_\mathbb{Q}}]^{-1}\},\\
\phi_{\mathcal{L}_{\mathbb{S}_\mathbb{Q}}}(v) &= [\Omega_{\mathcal{L}_{\mathbb{S}_\mathbb{Q}}}(v)]^{-1} -v^{-1}, \\
\varphi_{\mathcal{L}_{\mathbb{S}_\mathbb{Q}}}(\rho) &= \big(\rho + [\phi_{\mathcal{L}_{\mathbb{S}_\mathbb{Q}}}^{\mr{inv}}(\rho)]^{-1} \big)^{-1},
\end{align}\ES
where $\{\bf{A}_{\mathbb{S}_\mathbb{Q}}, \bf{x}_{\mathbb{S}_\mathbb{Q}}, \bf{n}_{\mathbb{S}_\mathbb{Q}}\}$ are the sub-matrix and sub-vectors corresponding to $\mathbb{S}_{\mathbb{Q}}$. Using Theorem~\ref{Them:Constrain_sum} and multi-user information theory\cite[Chapter 15]{InforTh}, the constrained capacity region of GMU-MIMO in the form of user group is given as follows.

\begin{theorem}[Constrained-Group-Capacity Region]\label{Pro:R_OAMP/VAMP}
	Assume that $\Omega_{\mathcal{S}}(\rho) = \varphi_{\mathcal{L}_{\mathbb{S}_\mathbb{Q}}}(\rho)$ has a unique positive solution $\rho^*$, and $v^*=\big([\Omega_{\mathcal{S}} (\rho^*)]^{-1} - \rho^*\big)^{-1}$.
	The constrained group capacity region $\{{\it{R}}_{\mathbb{S}_{\mathbb{Q}}}, { {\forall \, \mathbb{Q} \subseteq \mathbb{G}}}\}$ of GMU-MIMO in \eqref{Eqn:cons_gmu_recv} is
	 \BS\label{Eqn:multiu_cap}\begin{align}
      \!\!\!\!\!{\it{R}}_{\mathbb{S}_{\mathbb{Q}}}&\le I(\bf{A}  \bf{x}+\bf{n} ;\bf{x}_{\mathbb{S}_{\mathbb{Q}}}|\bf{x}_{\mathbb{S}_{\mathbb{Q}^c}} ) \\ 
     &= I(\bf{A}_{\mathbb{S}_{\mathbb{Q}}} \bf{x}_{\mathbb{S}_{\mathbb{Q}}} +\bf{n}_{\mathbb{S}_{\mathbb{Q}}} ;\bf{x}_{\mathbb{S}_{\mathbb{Q}}} ) \\
	&=\log \! \left| \bf{B}_{\mathbb{S}_\mathbb{Q}}\!(v^*) \right|+ N{\!_u}|\mathbb{S}_{\mathbb{Q}}| \Big( {\log} \Omega_{\mathcal{S}}(\rho^*)\! +\!\!\! \int_{0}^{\rho^*} \!\!\!\!{\Omega_{\mathcal{S}}(\rho) d\rho} \Big), 
	\end{align}\ES
	where $\bf{B}_{\mathbb{S}_\mathbb{Q}}(v) = v^{-1}\bf{I}_{\mathbb{S}_\mathbb{Q}} + { snr }\bf{A}_{\mathbb{S}_\mathbb{Q}}^{\rm{H}}\bf{A}_{\mathbb{S}_\mathbb{Q}}$ and $|\mathbb{S}_{\mathbb{Q}}|$ is the number of users in  $\mathbb{S}_{\mathbb{Q}}$.
\end{theorem}

Similarly, for Gaussian signaling, the constrained group capacity region in~\eqref{Eqn:multiu_cap} is degraded to the well-known Gaussian capacity region~\cite{tse2005fundamentals}. As a result, the conventional Gaussian capacity region is a special case of the proposed constrained region when the signaling is Gaussian.

\section{Theoretical Sum Capacity Optimality of MU-OAMP/VAMP}
In this section, we analyze the achievable sum rate of MU-OAMP/VAMP, which is equal to the constrained sum capacity of GMU-MIMO. Then, the optimal design principle of multi-user codes is given.

\subsection{Achievable Sum Rate and Constrained-Sum-Capacity Optimality of MU-OAMP/VAMP}
To verify the optimality of MU-OAMP/VAMP, we investigate the achievable sum rate of MU-OAMP/VAMP with error-free performance. 

\subsubsection{Code rate} The following lemma, proved in \cite{BhattadTIT2007}, shows the connection between the code rate and the MMSE decoding function. 

\begin{figure}[t]
    \centering
    \includegraphics[width=0.67\columnwidth]{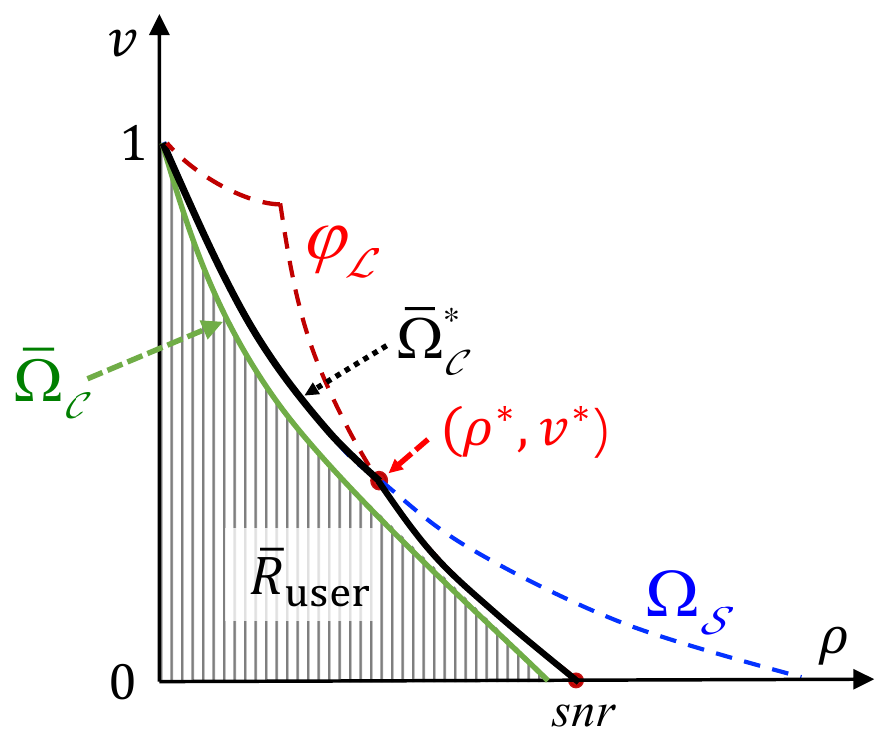}
	\caption{Illustration of the average user achievable rate of MU-OAMP/VAMP. ${\bar{\Omega}}_{\cal{C}}$ is upper bounded by ${\bar{\Omega}}_{\cal{C}}^*$ in~\eqref{Eqn:MU_opt_NLD}, i.e., the minimum of $\varphi_{\cal L}$ in~\eqref{Eqn:NewMUSE} and $\Omega_{\cal{S}}$ in~\eqref{Eqn:Omega_S}. The area covered by ${\bar{\Omega}}_{\cal{C}}$ equals the average rate of all users $\bar{R}_{\rm{user}}$.}\label{Fig:SVTF_area}
\end{figure}

\begin{lemma}[Code-Rate-MMSE]\label{Lem:rate-MMSE}
	The code rate $R_{\mathcal{C}_g}$ of $\mathcal{C}_g$ for the users in $\mathbb{S}_g$ is given by
	\BE\label{Eqn:Rg_mmse}
	R_{\mathcal{C}_g} =  \int_{0}^{\infty}\Omega_{{\cal{C}}_g}(\rho)d\rho.
	\EE
	Similarly, the average rate $\bar{R}_{\mr{user}}$ of all users is given by
	\BE\label{Eqn:Rsym_mmse}
	\bar{R}_{\mr{user}} =N_u \int_{0}^{\infty}{\bar{\Omega}}_{{\cal{C}}}(\rho)d\rho,
	\EE
	which is shown in Fig.~\ref{Fig:SVTF_area}. Actually, $\bar{R}_{\mr{user}}$ also denotes the user rate in a symmetric GMU-MIMO. 
\end{lemma}

\subsubsection{Upper bounds of decoding function} The following two lemmas provide two tight upper bounds of the MMSE decoding functions $\{\Omega_{\mathcal{C}_g}(\rho)\}$.

\begin{lemma}[Decoding Gain]\label{Pro:MU_Dec gain}The channel decoding can provide significant coding gains, such that the locally optimal decoding of APP decoders should do better than locally symbol-by-symbol demodulation,
	\BE\label{Eqn:coding_gain_uk}
	\Omega_{\mathcal{C}_g}(\rho)< \Omega_{\mathcal{S}}(\rho),\;\; \forall \rho \geq 0, \;\; \forall g.
	\EE
	Thus, 
	\BE\label{Eqn:coding_gain_Mu}
	{\bar{\Omega}}_{\mathcal{C}}(\rho)< \Omega_{\mathcal{S}}(\rho),\;\; \forall \rho \geq 0.
	\EE
\end{lemma}

According to (\ref{Eqn:MU-TF}) and (\ref{Eqn:NewMUSE}), Fig.~\ref{Fig:SVTF_area} shows that MU-OAMP/VAMP receiver is error-free if and only if the NLD curve ${\bar{\Omega}}_{\mathcal{C}}(\rho)$ lies below the LD $\varphi_{\mathcal{L}}(\rho)$\cite{LeiOptOAMP}. Therefore, we obtain the error-free condition of MU-OAMP/VAMP in Lemma~\ref{Pro:MU_error_free}.
\begin{lemma}[Error-Free Decoding]\label{Pro:MU_error_free}
	MU-OAMP/VAMP can achieve error-free decoding if and only if
	\BE
	{\bar{\Omega}}_{\mathcal{C}}(\rho)< \varphi_{\mathcal{L}}(\rho), \;\;\; \forall\rho \in[0,  snr ) 
	\EE	
	and ${\bar{\Omega}}_{\mathcal{C}}(\rho) = 0$ for $\rho \ge  snr $.
\end{lemma}

Therefore, as shown in Fig.~\ref{Fig:SVTF_area}, the MMSE of a feasible coded NLD is defined as
\BE\label{Eqn:MU_opt_NLD}
{\bar{\Omega}}_{\mathcal{C}}^*({\rho})= 
\min\{\Omega_{\mathcal{S}}(\rho),\; \varphi_{\mathcal{L}}(\rho)\},\;\;\;\forall\rho \in[0,  snr), 
\EE
and ${\bar{\Omega}}_{\mathcal{C}}^*({\rho}) = 0$ for $\rho \ge  snr $.

From Lemma~\ref{Pro:MU_Dec gain} and Lemma~\ref{Pro:MU_error_free}, it is easily obtained as
\BE\label{Eqn:MU-upper_bound}
{\bar{\Omega}}_{\cal{C}}({\rho}) < {\bar{\Omega}}_{\mathcal{C}}^*({\rho}),\;\;\;\forall\rho \in[0,  snr).
\EE

\subsubsection{Achievable sum rate and constrained-sum-capacity optimality} 
Assuming that there exist code-books $\{{\cal{C}}_g\}$ for $\{\mathbb{S}_g\}$, ${\bar{\Omega}}_{\mathcal{C}}({\rho})$ can match ${\bar{\Omega}}_{\mathcal{C}}^*({\rho})$, i.e., ${\bar{\Omega}}_{\mathcal{C}}({\rho})=\tfrac{1}{G}\sum_{g=1}^{G}\Omega_{{\cal{C}}_g}(\rho) \to {\bar{\Omega}}^*_{\cal{C}}({\rho})$. The achievable sum rate of MU-OAMP/VAMP is
\BS \label{Eqn:MU-A1}
\begin{align} 
&R^{\rm{sum}}_{\rm{MU-OAMP/VAMP}} \to \tfrac{N}{G} \textstyle\sum\nolimits_{g=1}^{G} R_{\mathcal{C}_g}=\!N\int_{0}^{snr} {\bar{\Omega}}_{\cal{C}}^*({\rho}) d\rho \\ 
&= N\bar{R}_{\rm MU-OAMP/VAMP}, 
\end{align}
\ES
where $\bar{R}_{\rm MU-OAMP/VAMP}\equiv\int_{0}^{snr} {\bar{\Omega}}_{\mathcal{C}}^*({\rho}) d\rho$ denotes the average rate per transmit antenna.  The following theorem shows that the achievable sum rate of MU-OAMP/VAMP  is  equal to the constrained sum capacity of GMU-MIMO.

\begin{theorem}[Constrained-Sum-Capacity Optimality]\label{Pro:MU_R_OAMP/VAMP}
	Assume that $\Omega_{\mathcal{S}}(\rho)=\varphi_{\mathcal{L}}(\rho)$ has a unique positive solution $\rho^*$ and $v^*=\big([\Omega_{\mathcal{S}} (\rho^*)]^{-1} - \rho^*\big)^{-1}$. Then, $\bar{R}_{\rm MU-OAMP/VAMP}=\bar{C}$ and
	\BE\label{Eqn:MU-achie_rate}
	R_{\rm{MU-OAMP/VAMP}}^{\rm{sum}} = C_{\mr{GMU-MIMO}}^{\mr{sum}}.
	\EE
\end{theorem}
\begin{IEEEproof} 
	See APPENDIX~\ref{APP:Cap_Opt}. 
\end{IEEEproof}

\begin{figure}[t]
\centering
\includegraphics[width=0.67\columnwidth]{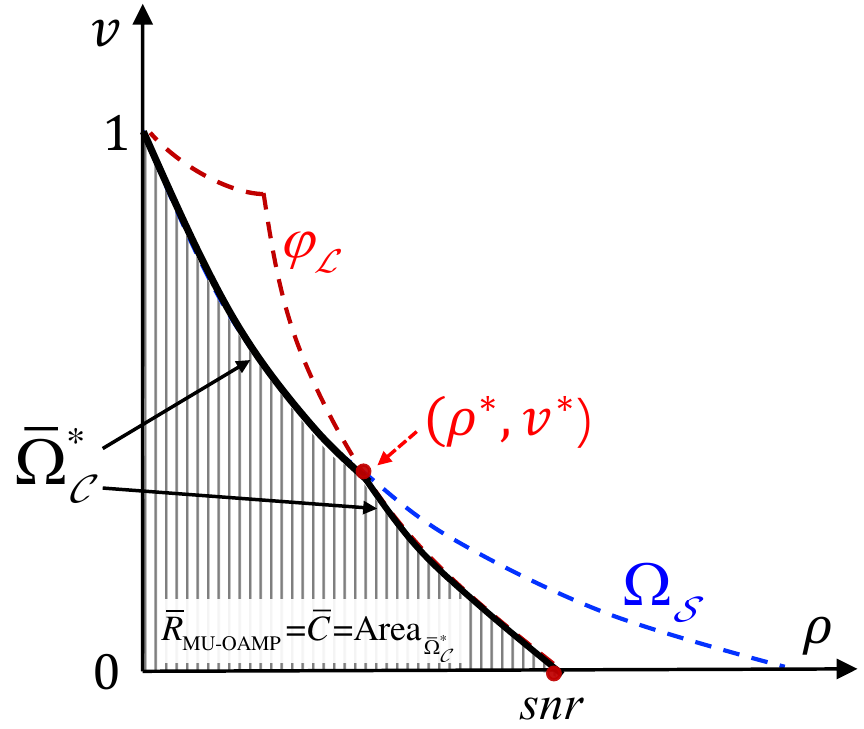}\vspace{-0.1cm}
\caption{Illustration of the average constrained capacity of GMU-MIMO, the average rate of MU-OAMP/VAMP, and the optimal MMSE function of decoder. The maximum achievable rate of MU-OAMP/VAMP equals the average constrained capacity, which is the area covered by ${\bar{\Omega}}_{\cal C}^*$.}\label{Fig:OAMP/VAMP_area}
\end{figure}

\begin{figure}[!t]
	\centering
	\includegraphics[width=1\columnwidth]{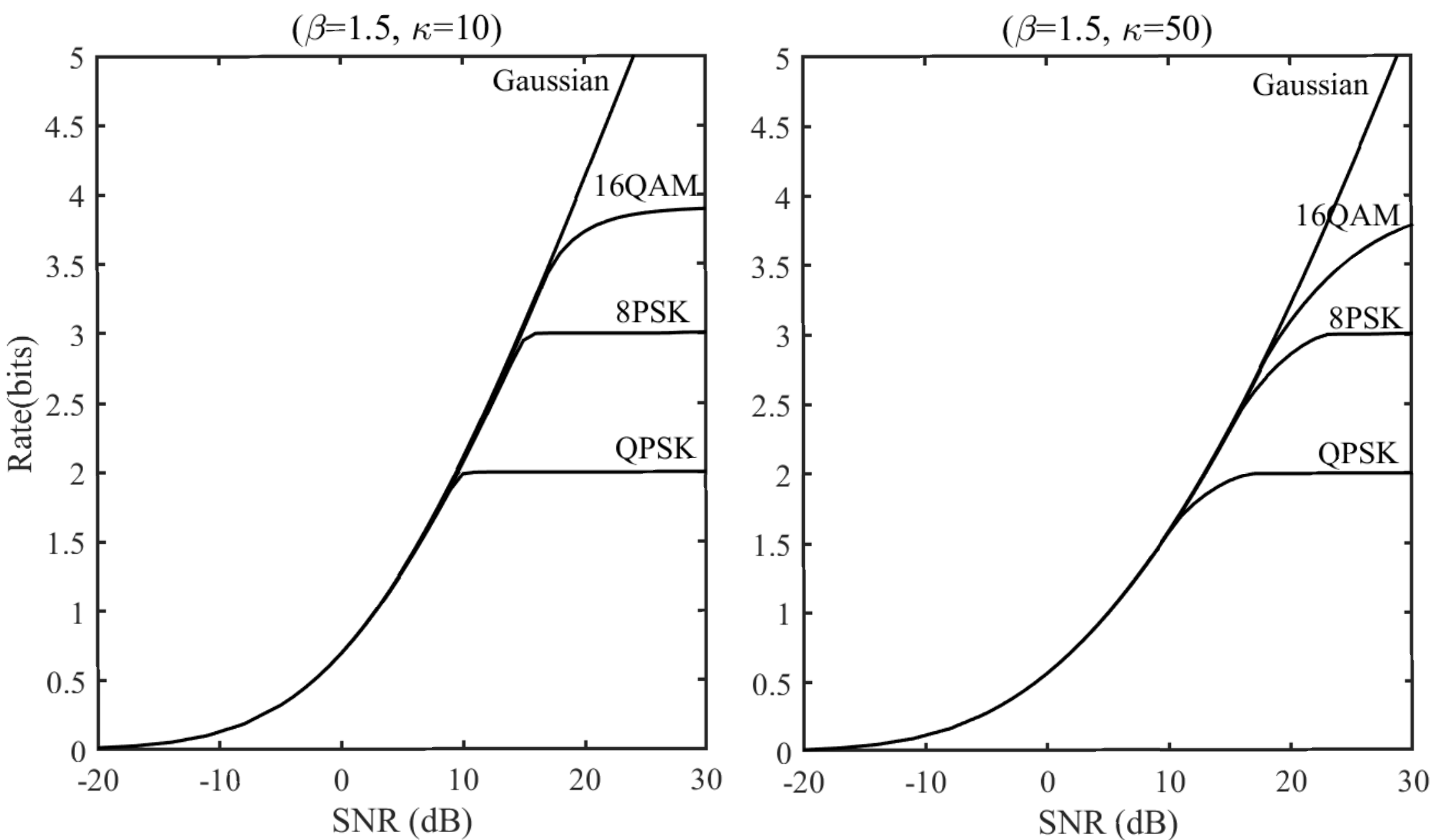}\vspace{-0.5cm}
	\caption{The average achievable rate $\bar{R}_{\rm MU-OAMP/VAMP}$ (per transmit antenna) of MU-OAMP/VAMP in the GMU-MIMO with $\beta=N/M=1.5$ and $\kappa=\{10,50\}$ when considering QPSK, 8PSK, 16QAM, and Gaussian signaling.}\label{Fig:sum_MU_OAMP/VAMP}
\end{figure}
As shown in Fig.~\ref{Fig:OAMP/VAMP_area}, from Theorem~\ref{Pro:MU_R_OAMP/VAMP}, \eqref{Eqn:const_SC}, and \eqref{Eqn:MU-A1}, the average constrained capacity of GMU-MIMO $\bar{C}$, the average rate of MU-OAMP/VAMP~${\bar{\it{R}}}_{\rm{MU-OAMP/VAMP}}$, and the area covered by the optimal MMSE function of decoder ${\bar{\Omega}}_{\cal{C}}^*(\rho)$ are equal. This also illustrates the constrained-sum-capacity optimality of MU-OAMP/VAMP.

Fig.~\ref{Fig:sum_MU_OAMP/VAMP} shows that the average achievable rate (per transmit antenna) of MU-OAMP/VAMP in GMU-MIMO system with  channel load $\beta=N/M=1.5$, conditional number of channel matrix $\kappa=\{10,50\}$, and QPSK, 8PSK, 16QAM and Gaussian signaling.  MU-OAMP/VAMP  achieves the Gaussian capacity for Gaussian signaling. For QSPK, 8PSK, and 16QAM modulations,  MU-OAMP/VAMP can achieve the sum constrained capacity under the unique fixed point assumption.

In summary,  we have demonstrated the information-theoretical (i.e., constrained-sum-capacity) optimality of MU-OAMP/VAMP over the coded GMU-MIMO with arbitrarily distributed signaling and general right-unitarily-invariant channel matrices. In the next subsection, we will further discuss the rate allocation for each user group based on Theorem \ref{Them:Constrain_sum} and Theorem \ref{Pro:MU_R_OAMP/VAMP}.

\subsection{Multi-User Code Design for MU-OAMP/VAMP}\label{Sec:Cap_Opt}
To maximize the achievable sum rate of MU-OAMP/VAMP, $\{\Omega_{\mathcal{C}_g}(\cdot)\}$ is elaborately selected  to make ${\bar{\Omega}}_{\mathcal{C}}({\rho})$ match with ${\bar{\Omega}}_{\mathcal{C}}^*({\rho})$. From~\eqref{Eqn:Rg_mmse}-\eqref{Eqn:MU-A1}, the constraints of $\{\Omega_{\mathcal{C}_g}(\cdot)\}$ are give in Property~\ref{Pro:Const_code}.
\begin{property}[Matching Conditions]\label{Pro:Const_code}
	The optimal  $\{\Omega_{\mathcal{C}_g}(\cdot)\}$ for MU-OAMP/VAMP  satisfies the following conditions.
	\begin{itemize}
		\item $\{\Omega_{\mathcal{C}_g}(\rho), \; \forall g\in \mathbb{G}\}$ are monotone decreasing in $\rho\geq0$.\vspace{0.2cm}
		\item ${\bar{\Omega}}_{\mathcal{C}}(\rho)=\tfrac{1}{G}\Sigma_{g=1}^{G}\Omega_{\mathcal{C}_g}(\rho)\to{\bar{\Omega}}_{\mathcal{C}}^*(\rho)$. \vspace{0.2cm}
		\item $0\leq\Omega_{\mathcal{C}_g}(\rho)<\Omega_{\mathcal{S}}(\rho), \;\: \forall g\in \mathbb{G}$.
	\end{itemize}
\end{property}

Based on Property~\ref{Pro:Const_code} and \eqref{Eqn:MU_opt_NLD}, we have 
\BE\label{Eqn:Omega_n}
\Omega_{\mathcal{C}_g}(\rho)= \left\{ \begin{array}{l}
	\Omega_{\mathcal{S}}(\rho), \qquad\quad\; 0\leq\rho<\rho^*\\
	\zeta_g\big(\varphi_{\mathcal{L}}(\rho)\big), \quad\;\; \rho^*\leq\rho\leq { snr } \\ 
\end{array} \right.\!\!,  \;\;\;\forall g\in \mathbb{G},
\EE
and $ \Omega_{\mathcal{C}_g}(\rho)=0, \rho>{ snr }$, where $\zeta_g(\cdot)$ is a variance allocation function.

\begin{lemma}[Sum Code Rate]\label{Lem:asym_sumrate}
	The sum rate of all users is 
	\BS\label{Eqn:asymRsum}
	\begin{align}
	R_{\mr{sum}}&={\tfrac{N}{G}\textstyle\sum_{g=1}^{G}R_{\mathcal{C}_g}}=N\int_{0}^{snr}{\bar{\Omega}}_{\mathcal{C}}^*({\rho}) d\rho\\
	&=N\bar{R}_{\rm MU-OAMP/VAMP},   
	\end{align}
	\ES
	where $R_{\mathcal{C}_g}$ is given in \eqref{Eqn:Rg_mmse}.
\end{lemma}

\subsubsection{Symmetric systems} Consider $\{\zeta_g(x)=x, g\in \mathbb{G}\}$, which corresponds to the symmetric case, i.e., all users have the same transfer function and rate, i.e.,
\BS
\begin{align}
\Omega_{\mathcal{C}_g}(\rho)&={\bar{\Omega}}_{\mathcal{C}}^*(\rho), \\
\bar{R}_{\mr{user}}&= \bar{R}_{\rm MU-OAMP/VAMP}.
\end{align}
\ES

\subsubsection{Group-asymmetric systems} In group-asymmetric systems,  the users in the same group  have the same rate but the users in the different groups  have different rates, which are determined by the variance allocation function $\{\zeta_g(\cdot)\}$. For simplicity, let $v={\bar{\Omega}}_{\mathcal{C}}^*(\rho)$ and $v_g=\Omega_{\mathcal{C}_g}(\rho)$. Meanwhile, we consider the following constraint of $v_g$ for $v \in[0, \Omega_\mathcal{S}(\rho^*)]$ similar as~\cite{LeiTSP2019}:
\BE\label{Eqn: g2}
\gamma_i (v_i^{-1} -  c^*)=\gamma_g (v_g^{-1}-c^*), \quad\forall i,g\in \mathbb{G},
\EE
where $c^*=[\Omega_\mathcal{S}(\rho^*)]^{-1}$ and $\gamma_i$, $\gamma_g$ $\in [0, \infty)$. Note that \eqref{Eqn: g2} can be rewritten as  
\BS\label{Eqn:v_i}
\begin{align}
v_i &= [c^* +\gamma_i^{-1}{\gamma_g}(v_g^{-1}-c^*)]^{-1} \\
&= (b_{ig}v_g^{-1}+c_{ig})^{-1},  
\end{align}
\ES
where $b_{ig}=\gamma_i^{-1}{\gamma_g}\geq0$ and $c_{ig}=(1-b_{ig})c^*$ are fixed. Since $v_g^{-1}-c^*>0$, $v_i$ is increasing with $\gamma_i$ while decreasing with $\{\gamma_g, k\neq i\}$. Thus, we obtain the variance allocation function:
\BS
\begin{align}
\zeta_g(v_g) &\equiv v\\
&=\tfrac{1}{G}\textstyle\sum\nolimits_{g=1}^G (b_{ig}v_g^{-1}+c_{ig})^{-1},
\end{align}
\ES
which is a monotone increasing function for $v_g$. Based on \eqref{Eqn:v_i}, $v_g=\zeta_g^{-1}(v)$ may not have a close form but it can be obtained numerically.  In addition, $\zeta_g(\cdot)$ is increasing with $\gamma_g$ and decreasing with $\{\gamma_i, i\neq g\}$.  Following \eqref{Eqn:Omega_n} and \eqref{Eqn:Rg_mmse}, we have the following lemma.

\begin{lemma}[Monotonicity]\label{Lem:asym_Rk}
	$\{R_{\mathcal{C}_g}\}$ is increasing with $\gamma_g$ while decreasing with $\{\gamma_i, i\neq g\}$.
\end{lemma}

Lemma~\ref{Lem:asym_Rk} provides an important guidance for the rate allocation, i.e., the code rate $\{R_{\mathcal{C}_g}\}$  can be changed flexibly by adjusting~$\{\gamma_g\}$, which is useful in the practical multi-user codes design.

\begin{figure}[!t]
	\centering
	\includegraphics[width=0.75\columnwidth]{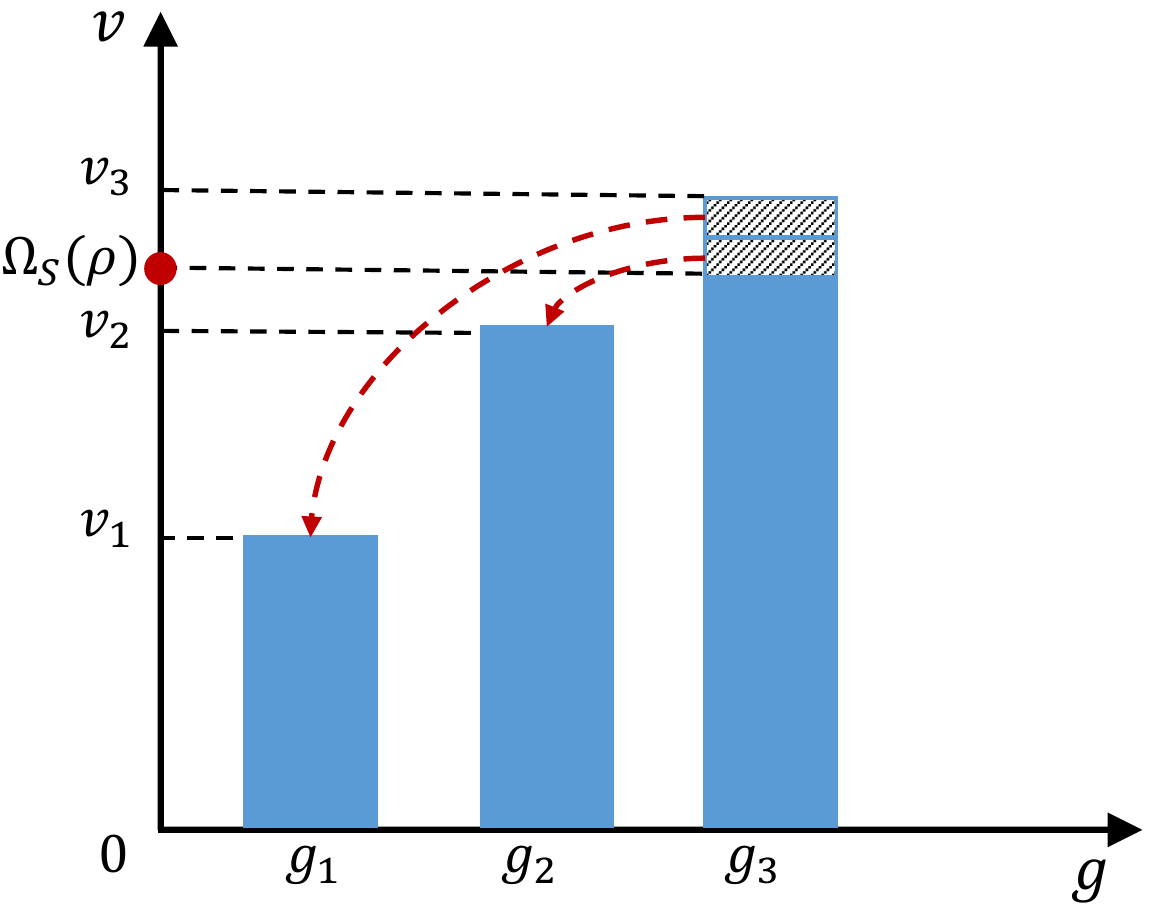}\vspace{-0.3cm}
	\caption{Adjustment of $\{v_g\}$ based on the constraints in \eqref{Eqn:cons_vk}.}\label{Fig:Rate_Allocate}
\end{figure}

\subsubsection{Transfer curves adjustment} 
Following Property~\ref{Pro:Const_code},  $\{v_g\}$ should satisfy  
\BS\label{Eqn:cons_vk}
\begin{align}\label{Eqn:cons_vk1}
v_g&=\mr{min}\{\zeta_g(v),\Omega_{\cal{S}}(\rho)\}, \quad \forall g\in \mathbb{G},\\ \label{Eqn:cons_vk2}
\tfrac{1}{G} \textstyle\sum\nolimits_{g=1}^G v_g &=\frac{1}{G}\textstyle\sum\nolimits_{i=1}^G (b_{ig}v_k^{-1}+c_{ig})^{-1} \nonumber \\
&= v, \quad \forall v \in (\Omega_{\mathcal{S}}(\rho^*), 1].
\end{align}
\ES

We take $G=3$ as an example to show the adjustment of $\{v_g\}$. As shown in Fig.~\ref{Fig:Rate_Allocate}, $v_3 > \Omega_{\mathcal{S}}(\rho)$  does not satisfy \eqref{Eqn:cons_vk}. Then, we uniformly allocate the exceeded part $v_3-\Omega_{\mathcal{S}}(\rho)$ to other under-loaded groups $\{v_1, v_2\}$. If some of the updated variances still exceed $\Omega_{\mathcal{S}}(\rho)$, the above adjustment is performed again. The adjustment stops when \eqref{Eqn:cons_vk} holds for $\{v_k\}$. Furthermore, for any given $G$, the adjustment of the transfer curves can be performed similarly.

\subsubsection{Extension to general configurations} \label{Sec:general_feasibility} 
Based on~\eqref{Eqn:cons_vk}, the LD is still available, but the NLD should be adjusted for different configurations of users and antennas. When each group includes a different number of users and each user has a different amount of antennas, the calculation of $v_g$ in~\eqref{Eqn:cons_vk2} should be modified by averaging the variances at all the antennas of users, i.e., 
\BE
v_g=\frac{1}{|\mathcal{S}_g|\textstyle\sum\limits_{i=1}^{|\mathcal{S}_g|}N_g^i}\textstyle\sum\limits_{i=1}^{|\mathcal{S}_g|}\textstyle\sum\limits_{n=1}^{N_g^i}v_g^{i,n},
\EE
where $g\in \mathbb{G}, i\in \mathcal{S}_g, n=1, .., N_g^i$, $v_g^{i,n}$ is the variances at the $n$-th antenna of user~$i$, $|\mathcal{S}_g|$ denotes the number of users in group $g$, and $N_g^i$ is the number of antennas for user $i$. As a result, the rate of user~$i$ in group~$g$ is 
\BE
    R_g^i = N_g^i R_{{\mathcal{C}}_{g}},
\EE
where $R_{{\mathcal{C}}_{g}}$ is given in \eqref{Eqn:Rg_mmse}. On the other hand, the sum rate of group~$g$ is 
\BE
 R_g = |\mathcal{S}_g|\textstyle\sum\limits_{i=1}^{|\mathcal{S}_g|}N_g^i R_{{\mathcal{C}}_{g}},
\EE
which is proportional to the number of users and antennas.

\subsection{Comparing with Conventional Turbo-LMMSE Receiver}\label{Any:Comping_Turbo}
In the following, we compare the average achievable rate per transmit antenna of MU-OAMP/VAMP and Turbo-LMMSE~\cite{LeiTSP2019,YuhaoTWC2018,XiaojunTIT2014}, which shows that Turbo-LMMSE is sub-optimal in the average achievable rate for non-Gaussian signaling. For simplicity, we  consider the comparison in symmetric systems.

The main difference between MU-OAMP/VAMP and Turbo-LMMSE is explained as follows. The LMMSE-LD and MMSE-NLD of Turbo-LMMSE are extrinsic that require independent input/output errors. MU-OAMP/VAMP only requires orthogonal input/output errors, which is generally less stringent than the independent requirement.  

Assume that the transfer functions of the detector and the decoder in Turbo-LMMSE are matched. The average achievable rate per transmit antenna of Turbo-LMMSE is given in \cite{XiaojunTIT2014}
\BE
R_{\mr{Turbo-LMMSE}}=\log|\mathcal{S}|-\int_0^{+\infty} \Omega_\mathcal{S}(\rho+\phi_{\cal L}(\Omega_\mathcal{S}(\rho))) d\rho.
\EE

\begin{figure}[!t]
	\centering
    \subfigure[Turbo-LMMSE: Extrinsic MMSE-NLD]{
    \begin{minipage}[t]{1\linewidth}\label{Fig:Turbo_LMMSE} 
    	\centering
    \includegraphics[width=0.75\linewidth]{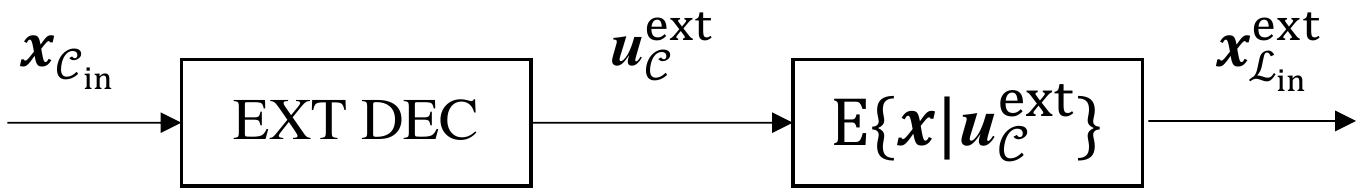}\vspace{0.3cm}
	\end{minipage}
	}\\
	\subfigure[MU-OAMP/VAMP: Orthogonal MMSE-NLD]
	{
    \begin{minipage}[t]{1\linewidth}\label{Fig:Turbo_OAMP/VAMP} 
    	\centering
    	\includegraphics[width=0.85\linewidth]{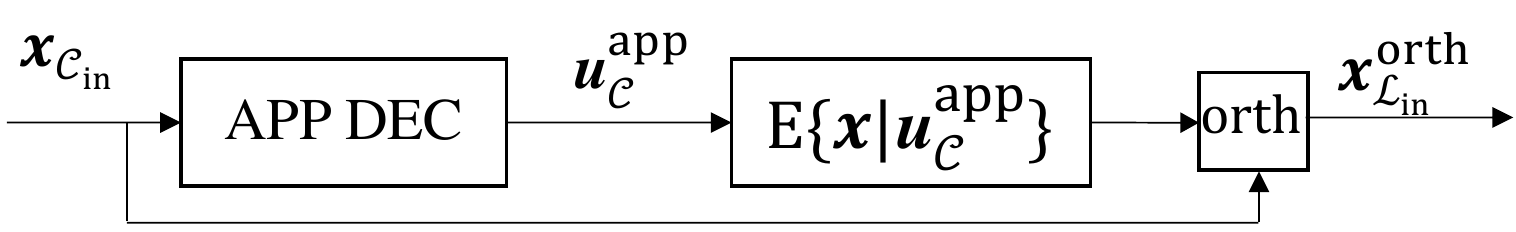}\vspace{0.3cm}
	\end{minipage}	
	}
	\caption{Comparison of Turbo-LMMSE and MU-OAMP/VAMP, where $\bf{x}_{\cal C_{\mr{in}}}$ is the input of NLD, Turbo-LMMSE uses extrinsic decoder (EXT DEC) and outputs extrinsic message $\bf{x}^{\mr{ext}}_{\cal L_{\mr{in}}}$, and MU-OAMP/VAMP uses APP decoder (APP-DEC) and outputs orthogonal message $\bf{x}^{\mr{orth}}_{\cal L_{\mr{in}}}$.}\label{Fig:OAMP/VAMP_Turbo}
\end{figure}

It rigorously proved that the MSE of MU-OAMP/VAMP is lower than that of Turbo-LMMSE in~\cite{MaTWC2019}. Here, we provide an intuitive explanation for this comparison from the perspective of the average achievable rate. Note that the LMMSE-LDs of MU-OAMP/VAMP and Turbo-LMMSE are equivalent for large-scale systems, in which the main difference lies in the NLD output operations. Fig. \ref{Fig:OAMP/VAMP_Turbo} shows the comparison between the MU-OAMP/VAMP NLD and Turbo-LMMSE NLD, where $\bf{x}_{\cal C_{\mr{in}}}$ is the input of NLD, $\bf{u}^{\mr{ext}}_{\cal C}$ (independent of $\bf{x}_{\cal C_{\mr{in}}}$) the extrinsic message of the decoder, and $\bf{u}_{\cal C}^{\mr{app}}$ the \emph{a-posteriori} message of the decoder. Then, the MMSEs of MU-OAMP/VAMP and Turbo-LMMSE are given by
\BS\begin{align}
\Omega_\mathcal{C}^{\mr{Turbo-LMMSE}}&=\mr{var}\{ \bf{x}| \bf{x}^{\mr{ext}}_{\cal L_{\mr{in}}}, \bf{x}_{\cal C_{\mr{in}}} \}  \nonumber \\
&= \mr{var}\{ \bf{x}| \mr{E}\{\bf{x}|\bf{u}^{\mr{ext}}_{\cal C}\}, \bf{x}_{\cal C_{\mr{in}}}\},\label{Eqn:Turbo_mmse}\\
\Omega_\mathcal{C}^{\mr{MU-OAMP/VAMP}}&=\mr{var}\{ \bf{x}| \bf{x}^{\mr{orth}}_{\cal L_{\mr{in}}}, \bf{x}_{\cal C_{\mr{in}}} \}\nonumber\\
&\mathop  = \limits^{(a)} \mr{var}\{ \bf{x}| \bf{u}^{\mr{app}}_{\cal C} \} \nonumber\\
&\mathop  = \limits^{(b)} \mr{var}\{ \bf{x}| \bf{u}^{\mr{ext}}_{\cal C}, \bf{x}_{\cal C_{\mr{in}}} \},\label{Eqn:OAMP/VAMP_mmse}
\end{align}\ES
where (a) follows orthogonal operation \eqref{Eqn:MU-OAMP/VAMP} that $\{\bf{x}^{\mr{orth}}_{\cal L_{\mr{in}}}, \bf{x}_{\cal C_{\mr{in}}}\}$ is a sufficient statistic for $\bf{u}^{\mr{app}}_{\cal C}$, and (b) follows that $\bf{u}^{\mr{app}}_{\cal C}$ is a sufficient statistic for $\{\bf{u}^{\mr{ext}}_{\cal C}, \bf{x}_{\cal C_{\mr{in}}}\}$. For Gaussian signaling, MU-OAMP/VAMP and Turbo-LMMSE are equivalent as $\mr{E}\{\bf{x}|\bf{u}^{\mr{ext}}_{\cal C}\}$ is a sufficient statistic of $\bf{u}^{\mr{ext}}_{\cal C}$. For non-Gaussian signaling, the expectation operation $\mr{E}\{\bf{x}|\bf{u}^{\mr{ext}}_{\cal C}\}$ may loss effective information. Thus, $ \Omega_\mathcal{C}^{\mr{Turbo-LMMSE}}$ $\geq$ $\Omega_\mathcal{C}^{\mr{MU-OAMP/VAMP}}$, i.e., MU-OAMP/VAMP has a lower MSE than Turbo-LMMSE. Then, following the Lemma~\ref{Lem:rate-MMSE}, we have the following lemma.
\begin{lemma}\label{lem:Turbo_LMMSE_OAMP/VAMP}
For Gaussian signaling, the average achievable rate per transmit antenna of MU-OAMP/VAMP is equal to that of Turbo-LMMSE, while for non-Gaussian signaling, the average achievable rate of MU-OAMP/VAMP is greater than or equal to that of Turbo-LMMSE.
\end{lemma}

\subsection{Design of Optimal MU-LDPC Codes}
Considering that LDPC code is a candidate coding scheme in future communication systems, we propose a kind of MU-LDPC code $\mathcal{C}_g(\lambda_g(X), \mu_g(X))$ with different rates for user group $\mathbb{S}_g$, where $g\in \mathbb{G}$, $\lambda_g(X)=\textstyle\sum_{d = 2}^{d_{v,\rm{max}}^g} {{\lambda_d^g}{X^{d - 1}}}$ and $\mu_g(X) = \textstyle\sum_{d = 2}^{d_{c,\rm{max}}^g} {{\mu_d^g}{X^{d - 1}}}$ are the respective degree distributions of variable and check nodes, and $d_{v,\rm{max}}^g$ and $d_{c,\rm{max}}^g$ are the corresponding maximum degrees of variable and check nodes. 

Based on property~\ref{Pro:Const_code} and \eqref{Eqn:MU_opt_NLD}, the optimal design goal of MU-LDPC codes of $G$ user groups is 
\BE\label{Eqn:codevk}
\tfrac{1}{G}\textstyle\sum_{g=1}^{G}\Omega_{{\cal{C}}_g}(\rho) ={\bar{\Omega}}_{\cal{C}}^*(\rho), \quad 0 \le \rho \le snr.
\EE
As a result, the MU-LDPC codes of $G$ user groups need to be optimized to reach \eqref{Eqn:codevk}, such that the sum rates can achieve the constrained sum capacity of GMU-MIMO. The detailed optimization of MU-LDPC codes is similar as~\cite[Section IV.A]{LeiTIT2021}, where $\{\Omega_{{\cal{C}}_g}(\rho),\forall g\in \mathbb{G}\}$ is determined by extrinsic information transfer (EXIT) analysis of MU-LDPC codes and the MMSE function of selected modulation type, i.e., $\Omega_{\mathcal{S}}(\rho)\equiv1-\frac{1}{\pi} \int\frac{\left|\sum_{l=1}^{|\mathcal{S}|}s_le^{-|y-\sqrt{\rho}s_l|^2}\right|^2} {|\mathcal{S}| \sum_{l=1}^{|\mathcal{S}|}e^{-|y-\sqrt{\rho}s_l|^2} } d y$ is the MMSE function for an arbitrary discrete constellation $\mathcal{S}=\{s_1,\cdots,s_{|\mathcal{S}|}\}$ with equal probability ${1}/{|\mathcal{S}|}$\cite{Lozano2006TIT}.

\begin{table*}[!t]\footnotesize
	\centering
	\caption{Complexity of the MU-OAMP/VAMP Receiver}\label{Complexity}
	\centering\setlength{\tabcolsep}{2.2mm}
	\begin{tabular}{|c|c|c|c|c|}
		\hline
		\multicolumn{2}{|c|}{Name}  & \multicolumn{3}{c|}{Complexity} \\ \hline
		\multicolumn{2}{|c|}{LMMSE detection}    & \multicolumn{3}{c|}{$\mathcal{O}(min\{MN^2+N^3,NM^2+M^3\}\tau_{\rm{max}})$}  \\ \hline
		\multicolumn{1}{|c|}{\multirow{4}{*}{\begin{tabular}[c]{@{}c@{}}MU-LDPC\\ decoders\end{tabular}}} & \multicolumn{1}{c|}{Demodulation} &\multicolumn{3}{c|}{$\mathcal{O}(K\log_2|\mathcal{S}|\tau_{\rm{max}})$}           \\ \cline{2-5} 
		\multicolumn{1}{|c|}{}     & \multirow{3}{*}{SPA decoding}     & \multicolumn{3}{c|}{$\mathcal{O}(K\tau_{\rm{max}})$}           \\ \cline{3-5} 
		\multicolumn{1}{|c|}{}     &                                   & \multicolumn{2}{c|}{$+/-$}     &  \multicolumn{1}{c|}{box-plus}   \\ \cline{3-5} 
		\multicolumn{1}{|c|}{}     &                                   & \multicolumn{2}{c|}{$(\frac{K}{G}\textstyle\sum_{g=1}^{G}\textstyle\sum_{d=2}^{d_{v,\rm{max}}^g}3N_v^g(d)d)\tau_{\rm{max}}$}     & \multicolumn{1}{c|}{$(\frac{K}{G}\textstyle\sum_{g=1}^{G}\textstyle\sum_{d=2}^{d_{c,\rm{max}}^g}N_c^g(d)d(d-2))\tau_{\rm{max}}$}     \\ \hline
		\multicolumn{2}{|c|}{MU-OAMP/VAMP}     & \multicolumn{3}{c|}{$\mathcal{O}((min\{MN^2+N^3,NM^2+M^3\}+K\log_2|\mathcal{S}|+K)\tau_{\rm{max}})$}       \\ \hline
	\end{tabular}
\end{table*}
\subsection{Complexity Analysis of MU-OAMP/VAMP Receiver}\label{Sec:Complex_Anlys} 
To verify practicability of the proposed system, we analyze the implementation complexity of MU-OAMP/VAMP receiver. Fig.~\ref{Fig:MU-OAMP/VAMP} shows that the MU-OAMP/VAMP receiver is consisted of a LMMSE detector and $G$ APP decoders. The complexity of the LMMSE detector is $\mathcal{O}(min\{MN^2+N^3,NM^2+M^3\}\tau_{\rm{max}})$\cite{YuhaoTWC2018}, where $\tau_{\rm{max}}$ is the maximum iteration number. The APP decoders include a bank of demodulation and LDPC decoding, which perform the symbol-by-symbol mapping (from estimation to likelihood probability) and sum-product algorithm (SPA)\cite{ryan2009channel} respectively. The complexity of demodulation is $\mathcal{O}(K\log_2|\mathcal{S}|\tau_{\rm{max}})$, where $K$ is the number of users and $|\mathcal{S}|$ is the size of modulation constellation $\mathcal{S}$. For MU-LDPC code $\mathcal{C}_g(\lambda_g(X), \mu_g(X))$, the number of variable nodes with degree $d$ and check nodes with degree $d$ are denoted by $N_v^g(d)$$=$$\left. N_uL\lambda_d^g / d \middle/\textstyle\sum_{d=1}^{d_{v,\rm{max}}^g} \lambda_d^g / d \right.$ and $N_c^g(d)=$ $\left. N_uL\mu_d^g / d \middle/ \textstyle\sum_{d=1}^{d_{c,\rm{max}}^g} \mu_d^g / d \right.$, which require $\textstyle\sum_{d=2}^{d_{v,\rm{max}}^g}3N_v^g(d)d$ additions/subtractions and $\textstyle\sum_{d=2}^{d_{c,\rm{max}}^g}N_c^g(d)d(d-2)$ box-plus operations~\cite{ryan2009channel} in one iterative detection. In summary, the complexity of the receiver is given in Table~\ref{Complexity}, in which the complexity of LMMSE detection is the decisive factor. 

Note that the complexity of the globally optimal \emph{maximum a posteriori} (MAP) receiver increases exponentially with the number of users and antennas\cite{verdu1984optimum}. As a result, compared with the MAP receiver, the MU-OAMP/VAMP receiver is acceptable in practical systems. Compared with Turbo-LMMSE \cite{LeiTSP2019,YuhaoTWC2018}, the incremental complexity of MU-OAMP/VAMP is ignorable. The main difference between MU-OAMP/VAMP and Turbo-LMMSE is the output operations of NLD. As shown in Fig.~\ref{Fig:OAMP/VAMP_Turbo}, MU-OAMP/VAMP requires additional orthogonal operations and \emph{a-posteriori} message of the decoder obtained by the sum of the extrinsic and \emph{a-prior} messages. 

Recently, a low-complexity memory AMP (MAMP) has been developed for right-unitarily-invariant matrices \cite{LeiMAMP}. Since MAMP uses a low-complexity memory-matched filter to suppress linear interference, its complexity is comparable to AMP and much lower than that of OAMP/VAMP.  Moreover, state evolution can accurately characterize the dynamics of MAMP. Aside from that, the state evolution of the MAMP reaches the MMSE fixed point predicted by the replication method. Therefore, to further achieve the lower complexity, MAMP\cite{LeiMAMP} is a good candidate for the proposed system.

\section{Two User-Group GMU-MIMO}\label{Sec:2userMIMO}
In this section, we take two user-group GMU-MIMO as
an example, i.e., all users are partitioned into two groups $\mathbb{S}_1$ and $\mathbb{S}_2$. The MU-OAMP/VAMP has been first proved to achieve the associated capacity region. Then, practical MU-LDPC codes are designed for two user-group GMU-MIMO.

\begin{figure*}[!t]
	\centering
	\includegraphics[width=0.85\linewidth]{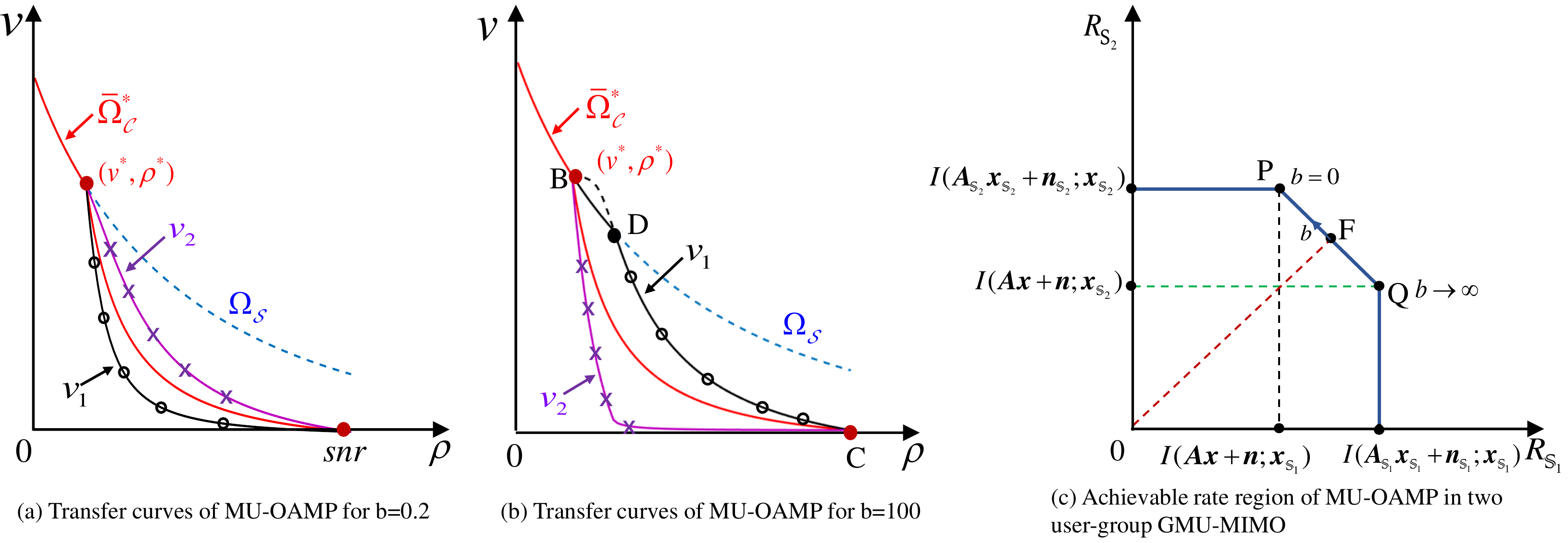}\vspace{-0.2cm}
	\caption{SE and achievable rate region of MU-OAMP/VAMP for two user-group GMU-MIMO systems.}\label{Fig:2AsymUser}
\end{figure*}

\subsection{Constrained-Capacity Optimality of MU-OAMP/VAMP in Two User-Group GMU-MIMO}
Let $v_1$ and $v_2$ denote the output NLD variances of  $\mathbb{S}_1$ and $\mathbb{S}_2$ respectively. Then 
\BE\label{Eqn:2userv1}
v_i=\left\{ \begin{array}{l}
	\Omega_{\mathcal{S}}(\rho), \qquad\; \;\; 0\leq\rho<\rho^*\\
	{\zeta}_i^{-1}\big(\varphi_{\mathcal{L}}(\rho)\big), \;\;  \rho^* <\rho\leq { snr }
\end{array} \right.\!\!\!, \;\forall i\in \{1, 2\}. 
\EE

Assuming that $\gamma_1 = 1$, $b \equiv b_{21}={\gamma_2}^{-1}$, and $c\equiv c_{21}=(1-b)c^{*}$ according to \eqref{Eqn:v_i}, we can rewrite $v_2$ as $v_2= (bv_1^{-1}+c)^{-1}$. Then, the average output variance of NLD is 
\BE\label{Eqn:2userAvg}
v = \frac{v_1+v_2}{2} = \frac{v_1+(bv_1^{-1}+c)^{-1}}{2}. 
\EE
Since $v={\bar{\Omega}}_{\mathcal{C}}^*({\rho})$ is fixed, we can find the desired $v_1$ by changing $b$.

As shown in Fig.~\ref{Fig:2AsymUser}(b), when the value of $b$ is too large, $v_1$ will be larger than $\Omega_{\mathcal{S}}(\rho)$ in certain values of $\rho$. Thus, the last condition in Property \ref{Pro:Const_code} may not hold. In this case,  we clip the exceeded part (the black dash line $\text{BD}$) of $v_1$,  and correspondingly allocate this part to $v_2$ (the purple solid line $\text{BC}$) to satisfy the average constraint~\eqref{Eqn:2userAvg}. By choosing the different values of $b$, MU-OAMP/VAMP can achieve the whole constrained capacity region of GMU-MIMO, i.e.,
\BS\begin{align}
&{R_{\mathbb{S}_1}} \le I(\bf{A}_{\mathbb{S}_1}\bf{x}_{\mathbb{S}_1}+\bf{n}_{\mathbb{S}_1};\bf{x}_{\mathbb{S}_1}),\\
&{R_{\mathbb{S}_2}} \le  I(\bf{A}_{\mathbb{S}_2}\bf{x}_{\mathbb{S}_2}+\bf{n}_{\mathbb{S}_2};\bf{x}_{\mathbb{S}_2}),\\
&R_{\rm{sum}}= {R_{\mathbb{S}_1}+R_{\mathbb{S}_2}}  \le  I(\bf{A}\bf{x}+\bf{n};\bf{x}).
\end{align} 
\ES
As shown in Fig.~\ref{Fig:2AsymUser}(c), point $\text{P}$, point $\text{Q}$, and segment $\text{PQ}$ denote the maximum ${R_{\mathbb{S}_1}}$, the maximum ${R_{\mathbb{S}_2}}$, and the maximum sum rate region, respectively, where point $\text{F}$ represents the symmetric system, i.e., ${R_{\mathbb{S}_1}}={R_{\mathbb{S}_2}}$. Meanwhile, different rate allocations can be achieved by adjusting $b$.

\subsection{Achieving the Maximal Extreme Points of Two User-Group GMU-MIMO}
As shown in Fig.~\ref{Fig:2AsymUser}(c), the achievable rate region of MU-OAMP/VAMP in two user-group GMU-MIMO is dominated by the convex combination of maximal extreme points $\text{P}$ and $\text{Q}$. To demonstrate the optimality of MU-OAMP/VAMP, we show all the maximal extreme points that can be achieved by MU-OAMP/VAMP. Due to the symmetry between $\text{P}$ and $\text{Q}$, for simplicity, we analyze the  transfer curves of each user group for point $\text{Q}$.

At point $\text{Q}$, the achievable sum rates of each user group are 
\begin{align}
R_{\mathbb{S}_1}&=  I(\bf{A}_{\mathbb{S}_1}\bf{x}_{\mathbb{S}_1}+\bf{n}_{\mathbb{S}_1};\bf{x}_{\mathbb{S}_1}),\\
R_{\mathbb{S}_2}&=  I(\bf{A}\bf{x}+\bf{n};\bf{x}_{\mathbb{S}_2}).  
\end{align}

For ${R_{\mathbb{S}_1}}$, the local estimation in~\eqref{Eqn:lmmse_Ax} is rewritten as 
 \BE
 f_{\mr{lmmse}}({\bf{s}}_{\mathbb{S}_1})\!\equiv\![{ snr }\bf{A}_{\mathbb{S}_1}^{\rm{H}}\bf{A}_{\mathbb{S}_1} +v_{{\mathbb{S}_1}}^{-1}\bf{I}_{\mathbb{S}_1}]^{-1}[{ snr }\bf{A}_{\mathbb{S}_1}^{\rm{H}}\bf{y}_{\mathbb{S}_1}+ v_{{\mathbb{S}_1}}^{-1}{\bf{s}}_{\mathbb{S}_1}],
 \EE
and transfer function is 
\BE 
\varphi_{\mathcal{L}_{\mathbb{S}_1}}(\rho) = \big(\rho + [\phi_{\mathcal{L}_{\mathbb{S}_1}}^{\mr{inv}}(\rho)]^{-1} \big)^{-1}.
\EE

\begin{figure}[!t]
	\centering
	\includegraphics[width=0.8\columnwidth]{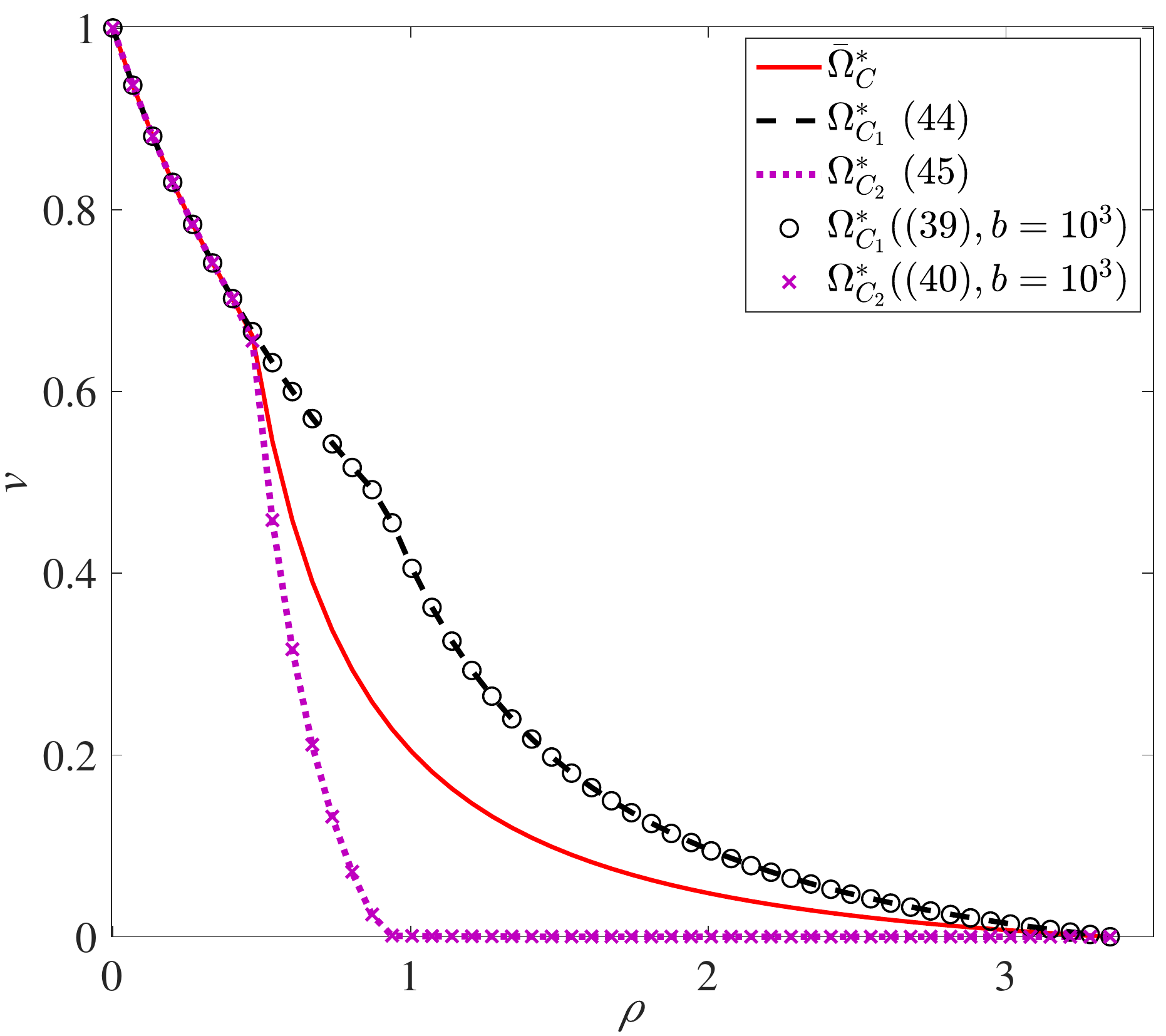}\vspace{-0.2cm}
	\caption{Achieving extreme point $\text{Q}$ by \eqref{Eqn:MU_opt_NLD_usr1}-\eqref{Eqn:MU_opt_NLD_usr2} and \eqref{Eqn:2userv1}-\eqref{Eqn:2userAvg} with $b=10^3$.}\label{Fig:max_point}
\end{figure}

Then, following~\eqref{Eqn:MU_opt_NLD}, the MMSE of a feasible coded NLD for user group~$1$ is 
\BE\label{Eqn:MU_opt_NLD_usr1}
{{\Omega}}_{\mathcal{C}_1}^*({\rho})= 
\min\{\Omega_{\mathcal{S}}(\rho),\; \varphi_{\mathcal{L}_{\mathbb{S}_1}}(\rho)\},\;\;\;\forall\rho \in[0,  snr), 
\EE
and ${{\Omega}}_{\mathcal{C}_1}^*({\rho}) = 0$ for $\rho \ge  snr $. Following~\eqref{Eqn:MU_opt_NLD}~and~\eqref{Eqn:2userAvg},
\BE\label{Eqn:MU_opt_NLD_usr2}
{{\Omega}}_{\mathcal{C}_2}^*=2*{\bar{\Omega}}_{\mathcal{C}}^*-{{\Omega}}_{\mathcal{C}_1}^*,
\EE
which is MMSE function of a feasible coded NLD for user group~$2$ with ${R_{\mathbb{S}_2}}$. As a result, the maximal extreme point $\text{Q}$ is achieved. Fig.~\ref{Fig:max_point} shows that transfer curves of ${{\Omega}}_{\mathcal{C}_1}^*$ and ${{\Omega}}_{\mathcal{C}_2}^*$ at the maximal extreme point $\text{Q}$ are obtained by \eqref{Eqn:MU_opt_NLD_usr1} and \eqref{Eqn:MU_opt_NLD_usr2}, which avoids adjusting parameter $b$ in~\eqref{Eqn:2userAvg}. It should be noted that the above strategy is consistent with the rate adjustment strategy discussed in \eqref{Eqn:2userv1} and \eqref{Eqn:2userAvg} as $b\to \infty$. 
\begin{figure*}[!t]
	\centering 
	\includegraphics[width=1\linewidth]{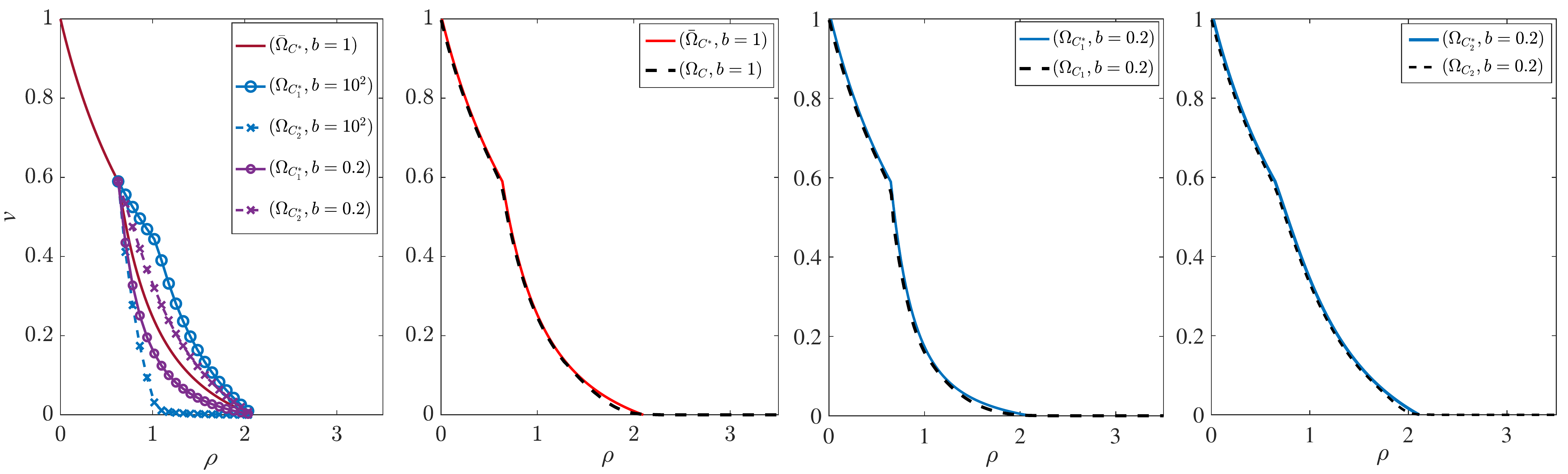}\vspace{-0.6cm}
	\caption{The SE curves of desired  ${\bar{\Omega}}_{{\cal{C}}}^*$, $\{{\Omega}_{{\cal{C}}_g}^*\}$, and designed $\{{\Omega}_{{\cal{C}}_g}\}$ in the GMU-MIMO systems, $g\in\{1,2\}$.}\label{Fig:SE_MU_LDPC_OAMP/VAMP}
\end{figure*}

\subsection{Design of MU-LDPC Codes in Two User-Group GMU-MIMO}
Similar as~\cite{LeiTIT2021}, the MU-LDPC codes are optimized for two user-group GMU-MIMO, where channel loads $\beta=1.5$ with $(N, M)=(500,333)$, and conditional number $\kappa=10$. We consider that there are $N/G=250$ antennas in each group.

For a target rate $\bar{R}_{\rm MU-OAMP/VAMP}=1$ in \eqref{Eqn:MU-A1}, i.e., $R_{\rm MU-OAMP/VAMP}^{\rm{sum}}=500$, with QPSK modulation, the limit SNR corresponding to the constrained sum capacity is $2.85$~dB given in Fig.~\ref{Fig:sum_MU_OAMP/VAMP}. As a result, the proposed MU-LDPC codes are given in Table~\ref{Opt_degree1} over the symmetric and group-asymmetric GMU-MIMO systems, where $b$ is the adjustment parameters in \eqref{Eqn:2userAvg} for rate allocation, user rate $R_{{\cal{C}}_g}$ of each group is equal to $N_uR_{\rm{LDPC}}\log_2|{\cal{S}}_{\rm{QPSK}}|$ with the rate of LDPC $R_{\rm{LDPC}}$, and QPSK modulation $|{\cal{S}}_{\rm{QPSK}}|=4$. Note that the decoding threshold of MU-LDPC codes is $( snr )^*_{\rm{dB}}=2.87$~dB and about $0.02$~dB away to the constrained sum capacity of GMU-MIMO.

To clearly visualize the above optimization process, Fig.~\ref{Fig:SE_MU_LDPC_OAMP/VAMP} provides the SE curves of desired ${\bar{\Omega}}_{\mathcal{C}}^*$ and $\{\Omega_{\mathcal{C}_g}^*\}$, and SE curves of designed $\{\Omega_{\mathcal{C}_g}\}$ for MU-LDPC codes over two user-group GMU-MIMO when considering the symmetric and asymmetric scenarios. In Fig.~\ref{Fig:SE_MU_LDPC_OAMP/VAMP}, the SE curve of desired $\Omega_{\mathcal{C}}^*$ is given for the symmetric case with $b=1$ and that of $\{{\Omega}_{\mathcal{C}_g}^*\}$ is given for the asymmetric cases with $b=100$ and $0.2$. Since  $\{\Omega_{{\cal{C}}_1}^*\}$ with $b=100$ is beyond the constraint of \eqref{Eqn:cons_vk1}, the SE curves of $\{\Omega_{{\cal{C}}_1}^*\}$ and $\{\Omega_{{\cal{C}}_2}^*\}$ with $b=100$ are adjusted according to \eqref{Eqn:cons_vk}. Note that the SE curves of designed MU-LDPC codes in Table~\ref{Opt_degree1} can match with those of desired MU-LDPC codes well, which also illustrates the optimality of the proposed MU-LDPC codes.

\section{Numerical Results}
In this section, we provide the practical finite-length bit-error rate (BER) performances of the proposed MU-LDPC codes for MU-OAMP/VAMP in GMU-MIMO. We benchmark the proposed GMU-MIMO framework with P2P-regular LDPC codes, well-designed P2P-irregular LDPC codes, and the state-of-the-art Turbo-LMMSE.

\subsection{Simulation Configuration}

\subsubsection{Signaling distribution} QPSK modulation is employed in simulations and the corresponding MMSE function $\Omega_{\mr{QPSK}}(\rho)$ is given in~\cite{LeiTIT2021}, where QPSK is a modulation scheme widely used in 5G and satellite communications and is also important for cell edge users and low-cost sensor devices,  i.e., relatively low SNR and low transmission rate scenarios. However, for higher-order modulations, e.g., $16$QAM, code design is very tough, because conventional code design such as EXIT analysis cannot be directly used due to three main reasons: 1) Probability density function (PDF) of the demodulated log-likelihood ratio (LLR) per bit in the constellation points is not Gaussian during the iterations; 2) PDFs of bit LLRs in the same constellation point are asymmetric and correlated; 3) Asymptotic decoding performance cannot be analyzed based on the transmission of all-zero codewords. To address the code design under higher-order modulations, bit-interleaved coded modulation
(BICM)  and superposition coded modulation (SCM) may be good candidates to approach perfect matching with the optimal code curve, in which BICM can be used to optimize constellation shaping for higher-order modulations during the iterations~\cite{Hanzo2009Twc} and SCM has been proven theoretically to asymptotically achieve perfect matching for Gaussian signals of interest~\cite{LeiTIT2021}. Hence, the design principle of multi-user codes and matching strategy are not limited to QPSK. Detailed discussions on the higher-order modulations are out of the scope of this paper, which is left as future work.

\subsubsection{User groups} For simplicity, we consider the symmetric and group-asymmetric GMU-MIMO with two user groups. Each group has $250$ transmitted antennas. The total number of transmitted antenna $N=500$. The coding length of each user in each group is $10^5$,  channel is used $400$ times for the whole transmission, and the corresponding parameters are given in Table~\ref{Opt_degree1}. Note that the user rate $R_{{\cal{C}}_g}$ of each group is equal to $N_uR_{\rm{LDPC}}\log_2|{\cal{S}}_{\rm{QPSK}}|$.

\subsubsection{Ill-Conditioned Channel Matrix} 
We assume that the channel is quasi-static and channel matrix $\bf{A}$ is fixed during the transmission, where $\bf{A} \in {\mathbb{C}}^{{\it{M}} \times {\it{N}}}$ is ill-conditioned, $N\!=\!500$, $M\!=\!333$, and channel load $\beta\!=\!N/M\!=\!1.5$. Let the SVD of $\bf{A}$ be $\bf{A}=\bf{U}\bf{\Lambda}\bf{V}$. $\bf{U}$ and $\bf{V}$ are generated by the orthogonal matrices in the QR decomposition of two IID Gaussian matrices. We set the eigenvalues $\{e_i\}$ in $\bf{\Lambda}$ as\cite{Vila2015ICASSP}: $e_i/e_{i+1} = \kappa^{1/\mathcal{T}}, i=1,...,\mathcal{T}-1$ and $\sum\nolimits_{i = 1}^{\mathcal{T}} {e_i^2 = N}$, where $\mathcal{T}\!=\!\text{min}\{M,N\}$ and $\kappa\!=\!\{10, 50\}$ denotes the condition number of $\bf{A}$. Furthermore, note that the advantages of OAMP/VAMP has been demonstrated in spatial correlated channels~\cite{MaTWC2019}, in which the channel matrices of $L$ channel uses is denoted as $\bf{A}=\mr{diag}\{\bf{A}_1, \bf{A}_2, ..., \bf{A}_L\}$, $\bf{A}_i=\bf{C}_{R}^{\frac{1}{2}}{\tilde{\bf{A}}}_i\bf{C}_{T}^{\frac{1}{2}}$, ${\tilde{\bf{A}}}_i$ is IID Gaussian matrix, and $\bf{C}_{R}^{\frac{1}{2}}$ and $\bf{C}_{T}^{\frac{1}{2}}$ are the receive and transmit correlation matrices. Meanwhile, the  spatial correlated channel matrix in\cite{MaTWC2019} is a special case of the ill-conditioned matrix. Therefore, for simplicity, only the ill-conditioned matrix is considered in this paper.

\subsection{BER Comparison with P2P Regular and Irregular LDPC Codes}\label{sec:P2P_REG-LDPC}
\begin{figure*}[!t]
	\centering
	\includegraphics[width=0.88\textwidth]{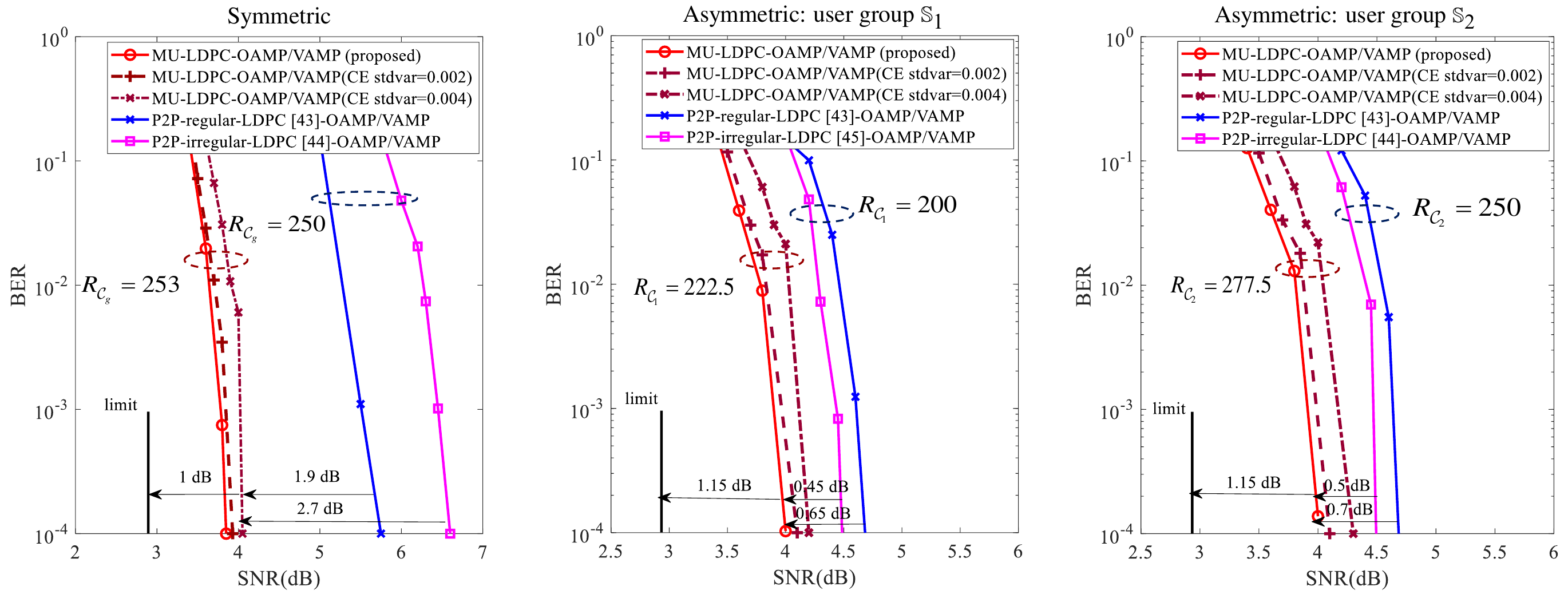}\vspace{-0.2cm}
	\caption{ 
		BER performances of GMU-MIMO with the proposed MU-LDPC codes given in Table~\ref{Opt_degree1}, the P2P-regular (3,6) LDPC codes with $R_{\mathcal{C}_g}=R_{\mathcal{C}_2}=250$ and (3,5) LDPC codes with $R_{\mathcal{C}_1}=200$~\cite{ryan2009channel}, and the P2P-irregular LDPC codes with $R_{\mathcal{C}_g}=R_{\mathcal{C}_2}=250$~\cite{Richardson2001} and $R_{\mathcal{C}_1}=200$~\cite{Kim2009}. $\beta=1.5$, $\kappa=10$, and $b=0.2$. In the imperfect channel estimations (CE) cases, the standard deviations (stdvar) of CE errors are $0.002$ and $0.004$.
	}\label{Fig:BER_OAMP/VAMP_Sym_Asym}
\end{figure*}

Fig.~\ref{Fig:BER_OAMP/VAMP_Sym_Asym} provides the BER simulations of the optimized MU-LDPC codes in Table~\ref{Opt_degree1} for GMU-MIMO systems. Meanwhile, in order to verify the advantages of the proposed frameworks, we also provide the BER performances of the P2P-regular LDPC codes and the well-designed P2P-irregular LDPC codes. The parameters of the P2P-regular LDPC codes are $(3,6)$ LDPC codes with coding rate $R_{\rm{LDPC}}=0.5$  and $(3,5)$ LDPC codes with coding rate $R_{\rm{LDPC}}=0.4$~\cite{ryan2009channel}. The degree distributions of one well-designed P2P-irregular LDPC code~\cite{Richardson2001} are $\lambda(X)=0.24426x+0.25907x^2+0.01054x^3+0.05510x^4+0.01455^7+0.01275x^9+0.40373x^{11}$ and $\mu(X)=0.25475x^6+0.73438x^7+0.01087x^8$, whose rate $R_{\rm{LDPC}}$ is 0.5 and the decoding threshold is 0.18~dB away from the P2P-AWGN capacity. 
The degree distributions of the other P2P-irregular LDPC code~\cite{Kim2009} are  $\lambda(X)=0.29472x+0.25667x^2+0.44861x^9$ and $\mu(X)=x^5$, whose rate $R_{\rm{LDPC}}$ is 0.4 and the decoding threshold is 0.18 dB away from the P2P-AWGN capacity.

\begin{table}[!t] \tiny \vspace{-0.2cm}
	\caption{proposed MU-LDPC codes with MU-OAMP/VAMP for tow user-group GMU-MIMO}\label{Opt_degree1}
	\centering\setlength{\tabcolsep}{0.1mm}{
		\begin{tabular}{|c||c|c|c|c|c|}
			\hline
			\multicolumn{1}{|c||}{\multirow {2}{*}{\tabincell{c}{\scriptsize System\\ \scriptsize parameters}}}&\multicolumn{1}{c|}{$\it{\beta}$}& \multicolumn{1}{c|}{$\it{\kappa}$}&\multicolumn{1}{c|}{$\textit{N}$} &\multicolumn{1}{c|}{$\textit{M}$} &\multicolumn{1}{c|}{\tabincell{c}{target \scriptsize$R_{\rm{sum}}$}}\\
			\cline{2-6}
			\multicolumn{1}{|c||}{} & \multicolumn{1}{c|}{1.5} & \multicolumn{1}{c|}{10} & \multicolumn{1}{c|}{500} & \multicolumn{1}{c|}{333} & \multicolumn{1}{c|}{500} \\
			\hline
			\multicolumn{1}{|c||}{\multirow {2}{*}{\scriptsize Scenarios}}& \multicolumn{1}{c|}{\multirow {1}{*}{\scriptsize symmetric}} & \multicolumn{4}{c|}{\scriptsize asymmetric}\\
			\cline{2-6}
			\multicolumn{1}{|c||}{} & \multicolumn{1}{c|}{$b={\text{1}}$} & \multicolumn{2}{c|}{$b={\text{100}}$} & \multicolumn{2}{c|}{$b={\text{0.2}}$} \\
			\hline
			\scriptsize $R_{\mathcal{C}_g}$ & 253 & 294 & 206 & 222.5 & 277.5\\
			\hline
			\scriptsize$R_{\rm{MU-OAMP/VAMP}}^{\rm{sum}}$ & 506 & \multicolumn{2}{c|}{500}& \multicolumn{2}{c|}{500}\\
			\hline
			$\mu(X)$&{\tabincell{c}{ ${\it{\mu}}_{\text{8}}=1$ }}&\multicolumn{2}{c|}{ ${\it{\mu}}_{\text{8}}=0.8$,  ${\it{\mu}}_{\text{12}}=0.2$}&\multicolumn{2}{c|}{ ${\it{\mu}}_{\text{8}}=0.8$, ${\it{\mu}}_{\text{25}}=0.2$}\\
			\hline
			\multicolumn{1}{|c||}{\multirow {10}{*}{$\lambda(X)$}} & $\lambda_{2}=0.4233$ & $\lambda_{2}=0.3571$& $\lambda_{2}=0.2465$& $\lambda_{2}=0.2824$& $\lambda_{2}=0.2692$\\
			\multicolumn{1}{|c||}{} & $\lambda_{3}=0.0677$    & $\lambda_{3}=0.2446$      & $\lambda_{3}=0.0221$        & $\lambda_{3}=0.0167$       & $\lambda_{3}=0.2631$    \\
			\multicolumn{1}{|c||}{} & $\lambda_{15}=0.0053$    & $\lambda_{11}=0.1689$      & $\lambda_{4}=0.1762$        & $\lambda_{7}=0.1376$       & $\lambda_{10}=0.0982$    \\  
			\multicolumn{1}{|c||}{} & $\lambda_{16}=0.2586$    & $\lambda_{12}=0.0186$      & $\lambda_{9}=0.1412$        & $\lambda_{8}=0.1738$       & $\lambda_{11}=0.0661$    \\    
			\multicolumn{1}{|c||}{} & $\lambda_{80}=0.1426$    & $\lambda_{30}=0.0383$      & $\lambda_{26}=0.0558$        & $\lambda_{35}=0.0766$       & $\lambda_{40}=0.0961$    \\  
			\multicolumn{1}{|c||}{} & $\lambda_{200}=0.1025$    & $\lambda_{35}=0.1725$      & $\lambda_{27}=0.1029$        & $\lambda_{40}=0.1233$       & $\lambda_{45}=0.0847$    \\ 
			\multicolumn{1}{|c||}{} & \multicolumn{1}{c|}{}    & \multicolumn{1}{c|}{}      & $\lambda_{70}=0.0300$        & $\lambda_{110}=0.0838$       & $\lambda_{250}=0.0663$    \\  
			\multicolumn{1}{|c||}{} & \multicolumn{1}{c|}{}    & \multicolumn{1}{c|}{}      & $\lambda_{80}=0.1190$         & $\lambda_{250}=0.1058$       & $\lambda_{300}=0.0563$    \\  
			\multicolumn{1}{|c||}{} & \multicolumn{1}{c|}{}    & \multicolumn{1}{c|}{}      & $\lambda_{250}=0.0710$         &\multicolumn{1}{|c|}{}       & \multicolumn{1}{|c|}{}    \\
			\multicolumn{1}{|c||}{} & \multicolumn{1}{c|}{}    & \multicolumn{1}{c|}{}      & $\lambda_{300}=0.0353$         & \multicolumn{1}{|c|}{}       & \multicolumn{1}{|c|}{}    \\
			\hline
			$(snr)^{\it{\ast}}_{\text{dB}}$ & \multicolumn{5}{c|}{2.87} \\
			\hline
			${\text{(limit)}_{\text{dB}}}$ & \multicolumn{5}{c|}{2.85}\\
			\hline
	\end{tabular}}\vspace{-0.2cm}
\end{table}
For symmetric GMU-MIMO, Fig.~\ref{Fig:BER_OAMP/VAMP_Sym_Asym} shows that the gap between BER curve at $10^{-4}$ of the optimized MU-LDPC code and the corresponding constrained sum capacity is $1$~dB, which verifies capacity-approaching performances of the optimized MU-LDPC codes. In addition, the optimized MU-LDPC codes  have about $1.9$~dB and $2.75$~dB performance gains over the benchmarks P2P-regular LDPC codes and the well-designed P2P-irregular LDPC codes, respectively. Meanwhile, user rate ($R_{\mathcal{C}_g}\!=\!253$) of the proposed MU-LDPC code is slightly higher than those of the benchmarks P2P-irregular and P2P-irregular LDPC codes ($R_{\mathcal{C}_g}\!=\!250$).

For user group $\mathbb{S}_1$ in Fig.~\ref{Fig:BER_OAMP/VAMP_Sym_Asym}, the optimized MU-LDPC codes with $R_{{\cal{C}}_1}=222.5$ can achieve about {$0.45$~dB and $0.65$~dB} performance gains over the P2P-irregular LDPC codes with $R_{{\cal{C}}_1}=200$, and the well-designed P2P-irregular LDPC codes with $R_{{\cal{C}}_1}=200$ respectively. For user group $\mathbb{S}_2$ in Fig.~\ref{Fig:BER_OAMP/VAMP_Sym_Asym}, the proposed MU-LDPC codes with $R_{{\cal{C}}_2}=277.5$ can achieve about {$0.5$~dB and $0.7$~dB} performance gains over the P2P-irregular and P2P-regular LDPC codes with $R_{{\cal{C}}_2}=250$, in which our codes also have the higher sum rate.  In short, comparing with the benchmarks, the proposed MU-LDPC codes not only has a significant improvement in BER performances, but also has higher transmission rates. The reason is that compared to P2P coding, multi-user coding suppress not only channel noise but also multi-user interference.

To confirm the robustness of the proposed MU-LDPC codes, we consider the BER simulations in GMU-MIMO with imperfect channel estimations, where the standard deviations of estimated channel errors are $0.002$ and $0.004$.  Fig.~\ref{Fig:BER_OAMP/VAMP_Sym_Asym} shows that imperfect channel estimations cause about $0.3$~dB performance losses, in which the gaps between BER curves at $10^{-4}$ of the proposed MU-LDPC codes and the corresponding theoretical limit are within $1.5$~dB. This verifies that the proposed MU-LDPC codes are robust to the imperfect channel estimations.

\subsection{Achieving the Whole Constrained-Capacity Region of Two User-Group GMU-MIMO}\label{sec:ach_max_sum}
\begin{figure}[!t]
	\centering
	\includegraphics[width=0.6\linewidth]{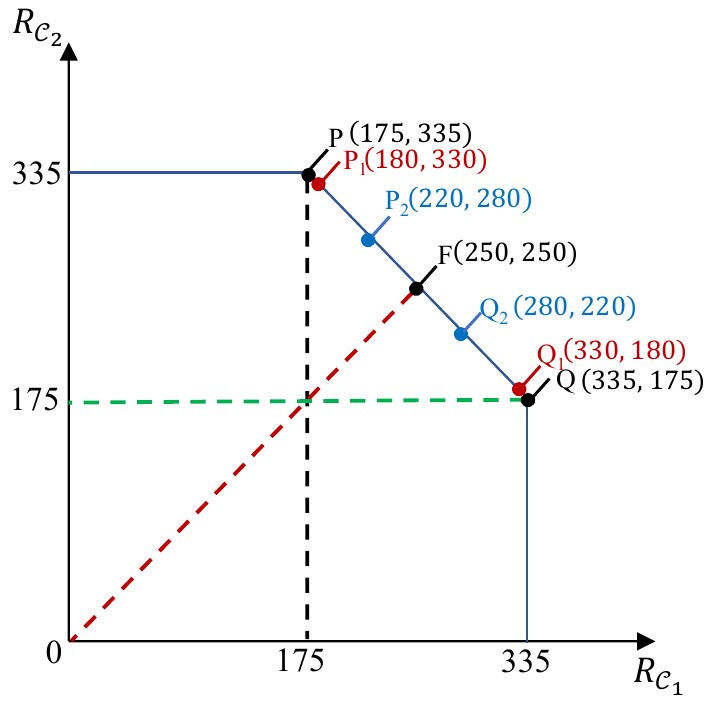}
	\caption{Extreme point $\text{Q}_{\text{1}}$, middle point $\text{Q}_{\text{2}}$, and symmetric point $\text{F}$ are achieved by the proposed MU-LDPC codes  in Table~\ref{Opt_degree2}. Due to the symmetry and adjustment of $b$, point $\text{P}_{\text{1}}$, point $\text{P}_{\text{2}}$, and any point at the region of sum capacity can be achieved by optimizing MU-LDPC codes.}\label{Fig:2user_maxrate} 
\end{figure}

To verify the optimality of the proposed frameworks, we consider the achievable of the sum capacity region of GMU-MIMO systems by optimizing MU-LDPC codes. Fig.~\ref{Fig:2user_maxrate} shows the sum capacity region of two user-group GMU-MIMO systems, where $\beta=1.5$, $\kappa=50$, and sum rate $R_{\text{MU-OAMP/VAMP}}^{\text{sum}}=R_{{\mathcal{C}}_{\text{1}}}+R_{{\mathcal{C}}_{\text{2}}}=510$. By setting the adjustment parameter $b \in \{10^3, 2.5, 1\}$ in \eqref{Eqn:2userAvg}, the MU-LDPC codes are optimized in Table~\ref{Opt_degree2} to achieve points $\text{Q}_{\text{1}}$, $\text{Q}_{\text{2}}$, and $\text{F}$ respectively. At point $\text{Q}_{\text{1}}$, the rate pair ($R_{{\mathcal{C}}_{\text{1}}}=330$, $R_{{\mathcal{C}}_{\text{2}}}=180$) corresponds to $(R_{\text{LDPC}_1}=0.66, R_{\text{LDPC}_2}=0.36)$, which closely approaches the rate pair ($R_{{\mathcal{C}}_{\text{1}}}=335$, $R_{{\mathcal{C}}_{\text{2}}}=175$) at theoretical   extreme point $Q$ with $(R_{\text{LDPC}_1}=0.67, R_{\text{LDPC}_2}=0.33)$. Note that the decoding threshold of MU-LDPC codes is $( snr )^*_{\rm{dB}}=5.25$~dB and about $0.02$~dB away to the sum capacity of GMU-MIMO, where the symmetric point $\text{F}$ and middle point $\text{Q}_{\text{2}}$ (achieved by the optimized MU-LDPC codes) approach the sum capacity. Due to the symmetry, extreme point $\text{P}_{\text{1}}$, middle point $\text{P}_{\text{2}}$ can be achieved in the same way by the MU-LDPC codes in Table~\ref{Opt_degree2} without re-optimizations. Therefore, the whole region of the maximal sum capacity of GMU-MIMO systems can be achieved by adjusting the parameter $b$ and optimizing the MU-LDPC codes.

Fig.~\ref{Fig:ber_maxpoint} provides the BER performances of the MU-LDPC codes with MU-OAMP/VAMP in Table~\ref{Opt_degree2} for two user-group GMU-MIMO systems, where $\beta\!=\!1.5$, $\kappa\!=\!50$, and $(snr)^*_{\rm{dB}}\!=\!5.25$~dB. For symmetric point~$\text{F}$, the MU-LDPC codes with $R_{\mathcal{C}_{g}}=250$ have about $3.9$~dB and $4.8$~dB performance gains over the benchmarks P2P-regular and P2P-irregular LDPC codes respectively.

At extreme point ${\text{Q}}_{\text{1}}$ and middle point ${\text{Q}}_{\text{2}}$, for user group $\mathbb{S}_1$, the optimized MU-LDPC codes with $R_{{\mathcal{C}}_{1}}=330\; ({\text{Q}}_{\text{1}})$ and $R_{{\mathcal{C}}_{1}}=280 \; ({\text{Q}}_{\text{2}})$ respectively have about $1.5$~dB and $1.7$~dB BER performance gains over the P2P-regular and P2P-irregular LDPC codes with $R_{{\mathcal{C}}_{1}}=250$.
For user group $\mathbb{S}_2$, the optimized MU-LDPC codes with $R_{{\mathcal{C}}_{2}}=180 \;({\text{Q}}_{\text{1}})$ and $R_{{\mathcal{C}}_{2}}=220 \;({\text{Q}}_{\text{2}})$ respectively have about $1.5$~dB and $1.7$~dB BER performance gains over the P2P-regular and P2P-irregular LDPC codes with $R_{{\mathcal{C}}_{1}}=200$.

\subsection{BER Comparison with Turbo-LMMSE}\label{sim:compare_Turbo}
In this subsection, we compare  the proposed framework with the state-of-the-art Turbo-LMMSE~\cite{LeiTSP2019,YuhaoTWC2018}. According to\cite[Section IV]{LeiTSP2019}, the transfer curves of asymmetric user groups in Turbo-LMMSE can be obtained based on extrinsic  LMMSE detector and extrinsic decoders. 

Based on extrinsic transfer curve matching, the MU-LDPC codes optimized for Turbo-LMMSE are given in Table~\ref{Opt_degree2}, in which the sum rates of symmetric and asymmetric cases are $500$, and the gap between the threshold of Turbo-LMMSE and the corresponding limit is $0.3$~dB. 

\begin{figure*}[!t]
	\centering
	\includegraphics[width=0.865\textwidth]{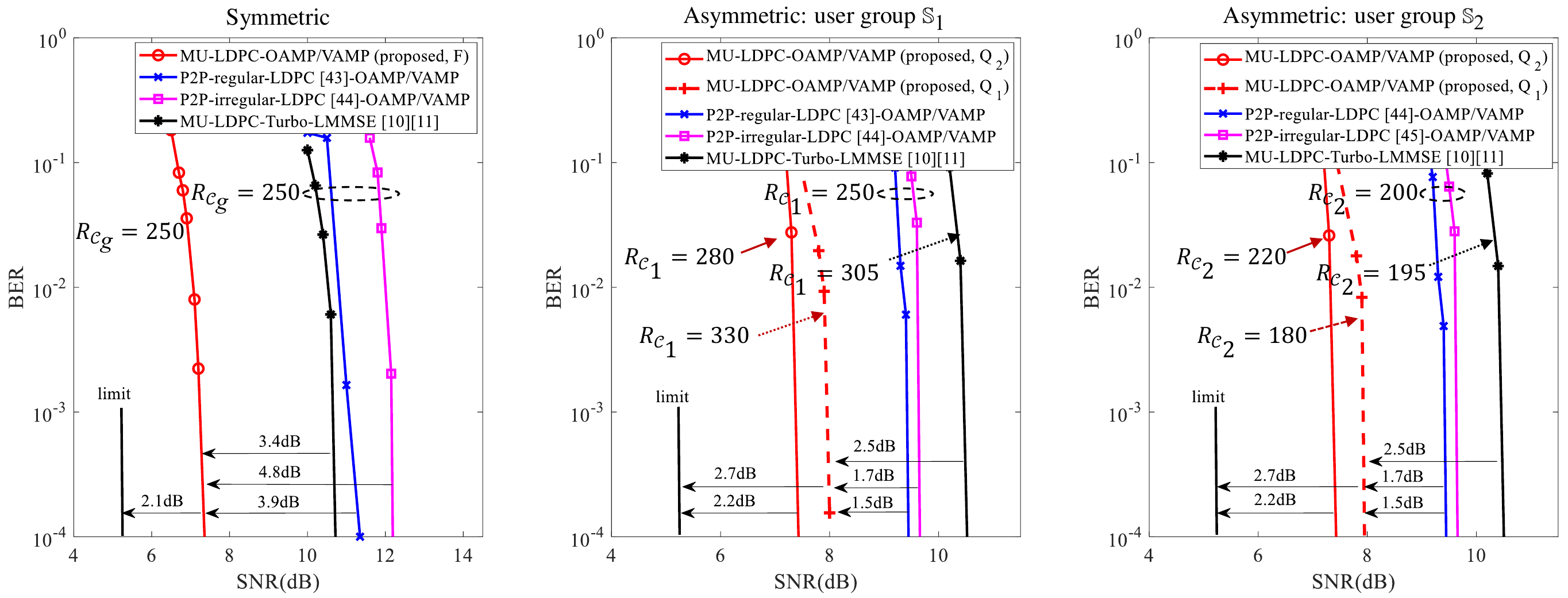}\vspace{-0.2cm}
	\caption{ BER performances of GMU-MIMO systems with the proposed MU-LDPC codes in Table~\ref{Opt_degree2}, the P2P-regular (3,6) LDPC codes with $R_{\mathcal{C}_g}=R_{\mathcal{C}_1}=250$, (3,5) LDPC codes with $R_{\mathcal{C}_2}=200$~\cite{ryan2009channel}, the P2P-irregular LDPC codes with $R_{\mathcal{C}_g}=R_{\mathcal{C}_1}=250$~\cite{Richardson2001} and $R_{\mathcal{C}_2}=200$~\cite{Kim2009}, and the MU-LDPC codes in Table~\ref{Opt_degree2} with Turbo-LMMSE~\cite{LeiTSP2019,YuhaoTWC2018}. $\beta=1.5$, $\kappa=50$, and $b \in \{1, 2.5, 10^3\}$.  
	}\label{Fig:ber_maxpoint}
\end{figure*}

As shown in Fig.~\ref{Fig:ber_maxpoint}, the proposed MU-LDPC code with MU-OAMP/VAMP has about $3.4$~dB performance gain over the Turbo-LMMSE in symmetric cases. For asymmetric cases, the proposed MU-LDPC codes at the extreme point $\text{Q}_1$ have  about $2.5$~dB performance gains.  Therefore, the above  comparisons verify the advantages of the proposed framework.

\begin{table*}[t] \tiny 
	\caption{proposed MU-LDPC codes designed for MU-OAMP/VAMP at Points $\text{F}$, $\text{Q}_{\text{1}}$, and $\text{Q}_{\text{2}}$ in Fig.~\ref{Fig:2user_maxrate} and MU-LDPC codes designed for Turbo-LMMSE~\cite{LeiTSP2019}.}\label{Opt_degree2}
	\centering\setlength{\tabcolsep}{0.25mm}{
		\begin{tabular}{|c||c|c|c|c|c|c|c|c|}
			\hline
			\multicolumn{1}{|c||}{\multirow {2}{*}{\tabincell{c}{\scriptsize System\vspace{-0.05cm}\\ \scriptsize parameters}}}&\multicolumn{1}{c|}{$\it{\beta}$}& \multicolumn{1}{c|}{$\it{\kappa}$}&\multicolumn{2}{c|}{$\textit{N}$} &\multicolumn{2}{c|}{$\textit{M}$} &\multicolumn{2}{c|}{\tabincell{c}{target \scriptsize$R_{\rm{sum}}$}}\\
			\cline{2-9}
			\multicolumn{1}{|c||}{} & \multicolumn{1}{c|}{1.5} & \multicolumn{1}{c|}{50} & \multicolumn{2}{c|}{500} & \multicolumn{2}{c|}{333} & \multicolumn{2}{c|}{510}\\
			\hline
			\multicolumn{1}{|c||}{\multirow {1}{*}{\scriptsize Methods}}&\multicolumn{5}{|c|}{\multirow {1}{*}{\scriptsize MU-OAMP/VAMP}}&\multicolumn{3}{|c|}{\multirow {1}{*}{\scriptsize Turbo-LMMSE}}\\
			\hline
			\multicolumn{1}{|c||}{\multirow {2}{*}{\scriptsize Scenarios}}& \multicolumn{1}{c|}{\multirow {1}{*}{\scriptsize symmetric}} & \multicolumn{4}{c|}{\scriptsize asymmetric}&\multicolumn{1}{c|}{\multirow {1}{*}{\scriptsize symmetric}} & \multicolumn{2}{c|}{\scriptsize asymmetric}\\
			\cline{2-9}
			\multicolumn{1}{|c||}{} & \multicolumn{1}{c|}{$\text{F}~(b=1)$} & \multicolumn{2}{c|}{$\text{Q}_{\text{1}}~(b=10^3)$} & \multicolumn{2}{c|}{$\text{Q}_{\text{2}}~(b=2.5)$} &\multicolumn{1}{c|}{$\text{F} ~(b=1)$} & \multicolumn{2}{c|}{$\text{Q}_{\text{1}}~(b=10^3)$}\\
			\hline
			\scriptsize $R_{\mathcal{C}_g}$ & 250 & 330 & 180 & 280 & 220 &250 & 305 & 195\\
			\hline
			\tiny$R_{\rm{MU-OAMP/VAMP}}^{\rm{sum}}$ & 500 & \multicolumn{2}{c|}{510}& \multicolumn{2}{c|}{500}& 500 & \multicolumn{2}{c|}{500}\\
			\hline
			\multicolumn{1}{|c||}{\multirow {1}{*}{$\mu(X)$}} & \multicolumn{8}{c|}{${\it{\mu}}_{\text{8}}=0.6$,~~~~${\it{\mu}}_{\text{25}}=0.4$}\\
			\hline
			\multicolumn{1}{|c||}{\multirow {10}{*}{$\lambda(X)$}} & $\lambda_{2}=0.3122$ & $\lambda_{2}=0.4671$& $\lambda_{2}=0.1175$ & $\lambda_{2}=0.3747$& $\lambda_{2}=0.2543$ & $\lambda_{2}=0.2912$ & $\lambda_{2}=0.4032$ & $\lambda_{2}=0.1340$ \\
			\multicolumn{1}{|c||}{} & $\lambda_{14}=0.0843$    & $\lambda_{3}=0.0473$      & $\lambda_{3}=0.1277$        & $\lambda_{18}=0.3103$       & $\lambda_{10}=0.3093$    & $\lambda_{10}=0.3014$    & $\lambda_{13}=0.2103$      & $\lambda_{4}=0.1650$  \\
			\multicolumn{1}{|c||}{} & $\lambda_{15}=0.2491$    & $\lambda_{20}=0.3665$      & $\lambda_{7}=0.0997$        & $\lambda_{19}=0.0044$       & $\lambda_{50}=0.0609$  & $\lambda_{50}=0.3057$    & $\lambda_{14}=0.1836$      & $\lambda_{5}=0.1032$   \\  
			\multicolumn{1}{|c||}{} & $\lambda_{110}=0.2089$    & $\lambda_{21}=0.0201$      & $\lambda_{8}=0.1569$        & $\lambda_{150}=0.2132$       & $\lambda_{60}=0.2074$  & $\lambda_{60}=0.0119$    & $\lambda_{70}=0.2029$      & $\lambda_{17}=0.0397$   \\    
			\multicolumn{1}{|c||}{} & $\lambda_{1000}=0.1455$    & $\lambda_{90}=0.099$      & $\lambda_{30}=0.0595$        & $\lambda_{200}=0.0056$       & $\lambda_{1000}=0.1681$  & $\lambda_{300}=0.0034$    & \multicolumn{1}{c|}{}      & $\lambda_{18}=0.3054$   \\
			\multicolumn{1}{|c||}{} & \multicolumn{1}{c|}{}    & \multicolumn{1}{c|}{}      & $\lambda_{35}=0.1239$        & $\lambda_{700}=0.0183$       & \multicolumn{1}{c|}{}  & $\lambda_{1000}=0.0864$    & \multicolumn{1}{c|}{}      & $\lambda_{90}=0.1428$   \\ 
			\multicolumn{1}{|c||}{} & \multicolumn{1}{c|}{}    & \multicolumn{1}{c|}{}      & $\lambda_{50}=0.0888$        & $\lambda_{800}=0.0735$       & \multicolumn{1}{c|}{} & \multicolumn{1}{c|}{}    & \multicolumn{1}{c|}{}      & $\lambda_{1000}=0.1099$\\  
			\multicolumn{1}{|c||}{} & \multicolumn{1}{c|}{}    & \multicolumn{1}{c|}{}      & $\lambda_{140}=0.0664$         & \multicolumn{1}{c|}{}      & \multicolumn{1}{c|}{} &\multicolumn{1}{|c|}{} & \multicolumn{1}{c|}{}    & \multicolumn{1}{c|}{}\\
			\multicolumn{1}{|c||}{} & \multicolumn{1}{c|}{}    & \multicolumn{1}{c|}{}      & $\lambda_{1000}=0.1596$         &\multicolumn{1}{|c|}{}       & \multicolumn{1}{|c|}{} & \multicolumn{1}{|c|}{} & \multicolumn{1}{c|}{}    & \multicolumn{1}{c|}{}  \\
			\hline
			$(snr)^{\it{\ast}}_{\text{dB}}$ & \multicolumn{5}{c|}{5.25} & \multicolumn{3}{c|}{7.42}  \\
			\hline
		    ${\text{(limit)}_{\text{dB}}}$ & \multicolumn{5}{c|}{5.23} &\multicolumn{3}{c|}{7.12} \\
			\hline
	\end{tabular}}
\end{table*}
\section{Conclusion}
This paper focuses on GMU-MIMO with the certain general and practical assumptions, i.e., practical channel coding, arbitrary discrete signaling, non-IID channel matrix, massive users and antennas, and available CSI only at the receiver. The information-theoretical limit of GMU-MIMO and the optimal capacity-achieving transceiver with practical complexity are open issues. To solve these issues, this paper proposes a unified framework for GMU-MIMO, jointly considering encoding, modulation, detection, and decoding. Meanwhile, group asymmetry is developed to make a tradeoff between user rate allocation and implementation complexity. Based on the framework, the constrained capacity region of group-asymmetric GMU-MIMO is accurately derived. The MU-OAMP/VAMP receiver with matched multi-user codes is proposed to achieve the error-free recovery performance for GMU-MIMO. The optimal design principle of multi-user code is presented and then a kind of MU-LDPC code is optimized for GMU-MIMO. 
Numerical results demonstrate that the gaps between theoretical decoding thresholds of the proposed framework with optimized MU-LDPC codes and the constrained sum capacity of GMU-MIMO are about 0.2 dB. Furthermore, their finite-length performances are  $1\sim2$~dB  away from the associated sum capacity. This implies that the proposed framework would be an important candidate for next-generation wireless communications.

\appendices
\section{Proof of Theorem~\ref{Pro:MU_R_OAMP/VAMP}}\label{APP:Cap_Opt}
Define $\phi_{\cal{L}}^{\mr{inv}}(\cdot)$ as the generalized inverse function of $\phi_{\cal{L}}(\cdot)$ as following.
\BE\nonumber
\phi_{\cal{L}}^{\mr{inv}}(\rho) =\left\{ \begin{array}{l}\vspace{-0.1cm}
		1, \qquad\quad\;\, \rho <\phi_{\cal{L}}(1),\\\vspace{-0.1cm}
		\phi_{\cal{L}}^{-1}(\rho), \quad\phi_{\cal{L}}(1)\leq\rho \leq { snr },\\
		0, \qquad\quad\;\, \rho >{ snr },
\end{array} \right.
\EE
where $\phi_{\cal{L}}^{-1}(\rho)$ is defined as the inverse function of $\phi_{\cal{L}}(\cdot)$ for $\rho\in[\phi_{\cal{L}}(1), { snr }]$. 

Then, we have
\BS\nonumber
\begin{align}
\varphi_{\cal{L}}(\rho) &\equiv \big(\rho + [\phi_{\cal{L}}^{\mr{inv}}(\rho)]^{-1} \big)^{-1}\\
&=\left\{   \begin{array}{l}
1/(\rho+1), \qquad\qquad\quad  \rho <\phi_{\cal{L}}(1), \vspace{-2mm}\\
\big(\rho + [\phi_{\cal{L}}^{-1}(\rho)]^{-1} \big)^{-1}, \quad\phi_{\cal{L}}(1)\leq\rho \leq { snr },\vspace{-2mm}\\ 
0, \qquad\qquad\qquad\qquad\;\; \rho >{ snr }.
\end{array} \right.
\end{align}
\ES

Define $\bar{b}(v) \equiv \tfrac{1}{N} \mathrm{Tr}\big\{[\bf{B}(v)]^{-1}\big\}$. Based on I-MMSE lemma~\cite{GuoTIT2005}, 
\begin{eqnarray}\nonumber
&\!\!\!\!\bar{R}_{\rm MU-OAMP/VAMP}&=\int_{0}^{\rho^*}\Omega_{\mathcal{S}}(\rho)d\rho + \int_{\rho^*}^{ snr}\varphi_{\mathcal{L}}(\rho) d\rho\nonumber\\
&\!\!\!\!\!\!\!\!\!\!\!\!\!\!\!\!\!\!\!\!\!\!\!\!\!\!\!\!\!\!\!\!\!\!\!\!\!\!\!\!\!\!\!\mathop = \limits^{(b)}&\!\!\!\!\!\!\!\!\!\!\!\!\!\!\!\!\!\!\!\!\!\!\!\!\!\!\!\!\! \!\!\!\!\!\!\!\!\!\!\!\! \int_{0}^{\rho^*} {\Omega_{\mathcal{S}}(\rho) d\rho} + \int_{\rho^*}^{\phi(0)}  {{\left[ {\rho } +  \left( \phi_{\mathcal{L}}^{\mr{inv}}{({\rho })}\right)^{ - 1} \right]}^{ - 1}d{\rho }}   \nonumber\\
&\!\!\!\!\!\!\!\!\!\!\!\!\!\!\!\!\!\!\!\!\!\!\!\!\!\!\!\!\!\!\!\!\!\!\!\!\!\!\!\!\!\!\!\mathop  = \limits^{(c)}&\!\!\!\!\!\!\!\!\!\!\!\!\!\!\!\!\!\!\!\!\!\!\!\!\!\!\!\!\!\!\!\!\!\!\!\! \!\!\!\!\!\int_{0}^{\rho^*} {\Omega_{\mathcal{S}}(\rho) d\rho} + \int_{v=v^*}^{v=0}  {\left( v^{-1} +  \phi_{\mathcal{L}}(v) \right)^{ - 1}d{\phi_{\mathcal{L}}(v) }}  \nonumber\\
&\!\!\!\!\!\!\!\!\!\!\!\!\!\!\!\!\!\!\!\!\!\!\!\!\!\!\!\!\!\!\!\!\!\!\!\!\!\!\!\!\!\!\!\mathop  = \limits^{(d)}&\!\!\!\!\!\!\!\!\!\!\!\!\!\!\!\!\!\!\!\!\!\!\!\!\!\!\!\!\!\!\!\!\!\!\!\!\! \!\!\!\!\!\int_{0}^{\rho^*} {\Omega_{\mathcal{S}}(\rho) d\rho}  - \!\!\!\int\limits_{v=v^*}^{v=0}\!\!{\Omega_{\mathcal{L}}(v^{-1})d{v^{-1} }} \!\!+\! \!\!\!\int\limits_{v=v^*}^{v=0} \!\!{ \Omega_{\mathcal{L}}(v^{-1}) d {\Omega_{\mathcal{L}}(v^{-1})}^{-1}} \nonumber\\
&\!\!\!\!\!\!\!\!\!\!\!\!\!\!\!\!\!\!\!\!\!\!\!\!\!\!\!\!\!\!\!\!\!\!\!\!\!\!\!\!\!\!\!\mathop  = \limits^{(e)}& \!\!\!\!\!\!\!\!\!\!\!\!\!\!\!\!\!\!\!\!\!\!\!\!\!\!\!\!\!\!\!\!\!\!\!\!\!\!\!\!\! \int_{0}^{\rho^*} {\Omega_{\mathcal{S}}(\rho) d\rho} - \!\! \int_{v=v^*}^{v=0}  \!\!{  \bar{b}(v) d{v^{-1} }} - \big[\log \bar{b}(v)\big]_{v=v^*}^{v=0}   \nonumber\\
&\!\!\!\!\!\!\!\!\!\!\!\!\!\!\!\!\!\!\!\!\!\!\!\!\!\!\!\!\!\!\!\!\!\!\!\!\!\!\!\!\!\!\!\mathop  = \limits^{(f)}&\!\!\!\!\!\!\!\!\!\!\!\!\!\!\!\!\!\!\!\!\!\!\!\!\!\!\!\!\!\!\!\!\!\!\!\!\!\!\!\!\!\!\! \int_{0}^{\rho^*} \!\! {\Omega_{\mathcal{S}}(\rho) d\rho} + \left[{ \tfrac{1}{N} \log \left| \bf{B}(v)  \right|}\right]_{v=0}^{v=v^*}- \big[\! \log  \bar{b}(v) \big]_{v=v^*}^{v=0} \nonumber\\
&\!\!\!\!\!\!\!\!\!\!\!\!\!\!\!\!\!\!\!\!\!\!\!\!\!\!\!\!\!\!\!\!\!\!\!\!\!\!\!\!\!\!\!\mathop  = \limits^{(g)}& \!\!\!\!\!\!\!\!\!\!\!\!\!\!\!\!\!\!\!\!\!\!\!\!\!\!\!\!\!\!\!\!\! \!\!\!\!\!\int_{0}^{\rho^*} \!\!\!\!{\Omega_{\mathcal{S}}(\rho) d\rho} + \!{ \tfrac{1}{N} \log \left| \bf{B}(v^*) \right|} \! +\!  \log  \bar{b}(v)\nonumber\\
&\!\!\!\!\!\!\!\!\!\!\!\!\!\!\!\!\!\!\!\!\!\!\!\!\!\!\!\!\!\!\!\!\!\!\!\!\!\!\!\!\!\!\!\mathop  = \limits^{(h)}&\!\!\!\!\!\!\!\!\!\!\!\!\!\!\!\!\!\!\!\!\!\!\!\! \!\!\!\!\!\!\!\!\!\!\!\! \log  \Omega_{\mathcal{S}}(\rho^*) + \int_{0}^{\rho^*} \!\!{\Omega_{\mathcal{S}}(\rho) d\rho} + \!{ \tfrac{1}{N} \log \left| \bf{B}(v^*) \right|} .\nonumber
\end{eqnarray}
Inequality $(a)$ is derived by the matching assumption and the equality holds if and only if there exists that code whose transfer function matches the $\bar{\Omega}_{C}^*(\rho)$. Equation $(b)$ follows the match assumption,  equation $(c)$ is based on the law $\int {\mathrm{Tr}\{ (s\bf{I} + \bf{A})^{-1}\} ds }$ $= \log\det (s\bf{I} + \bf{A})$, and equation $(d)$ is derived by $\det(\bf{I}_K+\bf{A}_{K\times M} \bf{B}_{M\times K}) = \det(\bf{I}_M+\bf{B}_{M\times K}\bf{A}_{K\times M})$ for any matrices $\bf{A}_{K\times M}$ and $\bf{B}_{M\times K}$. Equation $(e)$ follows the fix-point function $\tfrac{1}{|{\mathbb{S}_\mathbb{Q}}|} \mathrm{Tr}\big\{[\bf{B}_{{\mathbb{S}_\mathbb{Q}}}(v^*)]^{-1}\big\} =\Omega_{\mathcal{L}_{{\mathbb{S}_\mathbb{Q}}}}(v^*)= \Omega_{{\mathbb{S}_\mathbb{Q}}}(\rho^*)$. 
Therefore, from Theorem~\ref{Them:Constrain_sum} and  \eqref{Eqn:const_SC}, we can obtain $\bar{R}_{\rm MU-OAMP/VAMP}=\bar{C}$ and $R_{\rm{MU-OAMP/VAMP}}^{\rm{sum}}$ $=C_{\rm{MU-OAMP/VAMP}}^{\rm{sum}}$. 
\bibliographystyle{IEEEtran}
\bibliography{manuscript}
\end{document}